\documentclass[fleqn,usenatbib]{mnras}

\usepackage{newtxtext,newtxmath}
\usepackage{soul}
\usepackage[T1]{fontenc}
\DeclareRobustCommand{\VAN}[3]{#2}
\let\VANthebibliography\thebibliography
\def\thebibliography{\DeclareRobustCommand{\VAN}[3]{##3}\VANthebibliography}

\usepackage[dvipsnames]{xcolor}

\usepackage{graphicx}
\usepackage{amsmath}
\usepackage{amssymb}
\usepackage{xcolor}
\usepackage{changes}
\usepackage{adjustbox}
\usepackage{hyperref}

\title[Observational constraints on fraction of GWs from AGN]
{The most luminous AGN do not produce the majority of the detected
stellar-mass black hole binary mergers in the local Universe}

\author[N. Veronesi et al.]{
Niccolò Veronesi,$^{1}$\thanks{E-mail: veronesi@strw.leidenuniv.nl}
Elena Maria Rossi,$^{1}$
Sjoert van Velzen,$^{1}$
\\
$^{1}$ Leiden Observatory, Leiden University, PO Box 9513, 2300 RA Leiden,
The Netherlands}

\date{Accepted 2023 October 12. Received 2023 September 15;
in original form 2023 June 19}

\pubyear{2023}

\begin{document}
\label{firstpage}
\pagerange{\pageref{firstpage}--\pageref{lastpage}}
\maketitle


\begin{abstract}
Despite the increasing number of Gravitational Wave (GW) detections, the
astrophysical origin of Binary Black Hole (BBH) mergers remains elusive.
A promising formation channel for BBHs is inside accretion
discs around supermassive black holes, that power Active Galactic Nuclei (AGN).
In this paper, we test for the first time the spatial correlation between
observed GW events and AGN. To this end, we assemble all sky catalogues
with 1,412 (242) AGN with a bolometric luminosity greater than $10^{45.5}
{\rm erg\ s}^{-1}$ ($10^{46}\,{\rm erg\,s}^{-1}$) with spectroscopic redshift
of $z\leq0.3$ from the Milliquas catalogue, version 7.7b. These AGN are
cross-matched with localisation volumes of BBH mergers observed in the same
redshift range by the LIGO and Virgo interferometers during their first three
observing runs. We find that the fraction of the detected mergers originated
in AGN brighter than $10^{45.5}\,{\rm erg\,s}^{-1}$ ($10^{46}\,{\rm erg\,s}^{-1}$)
cannot be higher than $0.49$ ($0.17$) at a 95 per cent credibility level.
Our upper limits imply a limited BBH merger production efficiency of
the brightest AGN, while most or all GW events may still come from lower
luminosity ones. Alternatively, the AGN formation path for merging stellar-mass
BBHs may be actually overall subdominant in the local Universe. To our knowledge,
ours are the first observational constraints on the fractional contribution
of the AGN channel to the observed BBH mergers.
\end{abstract}



\section{Introduction}
\label{sec:intro}

The astrophysical mass spectrum of stellar-mass Black Holes (sMBHs) inferred
from the results of the first three observing runs of Advanced LIGO \citep{ligo15}
and Advanced Virgo \citep{acernese15} extends also to masses between $50\,
{\rm M}_\odot$ and  $120\,{\rm M}_\odot$ \citep{ligo21pop}.
This evidence challenges our current understanding of stellar evolution,
since no remnant with a mass in that range is expected to be the final stage
of the life of a single star \citep{heger02,belczynski16}. Pair Instability
Supernovae are expected to happen in that mass range, and are expected to
leave no compact remnant, thus opening a gap in the black hole mass spectrum
\citep{woosley19,mapelli21}.

The detection of mergers of sMBHs within this mass gap can be interpreted
as an evidence of binary formation channels beyond the “isolated stellar
binary” channel \citep[however, see also][]{demink16,costa21,tanikawa21}.
Other channels for Black Hole Binary (BBH) formation and merger involve
dense dynamical environments, such as Globular Clusters \citep{rodriguez16,
rodriguez18,rodriguez21}, Nuclear Star Clusters  \citep{antonini19,kritos22},
and accretion discs around Supermassive Black Holes (SMBHs) in Active Galactic
Nuclei (AGN) \citep{stone17,fabj20,ford22,mckernan22,li22b,li22,rowan22}.
The formation of binaries with massive components in all these dense environments
is facilitated by dynamical interactions such as exchanges in the case of
three-body encounters. In the interaction between a binary system and a third
object, the least massive of the three objects is expected to be scattered away
from the binary system, that is tightened by this process \citep{hills80,ziosi14}.
In case the gravitational potential of the host environment is deep enough to
retain the remnant of a BBH merger despite the post-merger recoil kick,
this can take part in a subsequent merger \citep{gerosa19}.
Binaries that merge in this so-called hierarchical scenario \citep{yang19,barrera22}
are expected to show specific signatures in the mass and spin distributions
of their components. Examples of these features are a low mass ratio, and
isotropically oriented spins \citep{gerosa17,gerosa21,tagawa21,wang21,
fishbach22,lilin22,mahapatra22}.

What differentiates AGN from other dynamically dense potential hosts of BBH mergers,
is the presence of a gaseous disc. Accretion discs around SMBHs are expected
to contain compact objects \citep{mckernan12b,tagawa20}.
The dynamical evolution of these objects is heavily influenced by the interaction
with the gas of the disc. This interaction is expected to make the sMBHs migrate
towards the innermost region of the AGN disk on timescales inversely proportional
to their mass \citep{mckernan12,delaurentiis22}. This migration should end
when the net torque exerted by the gas on the migrating compact object is
null. This is expected to happen at specific distances from the central SMBH,
the so-called “migration traps” \citep{bellovary16,peng21,grishin23}.

Due to the large localisation volumes associated to GW detections, the
fractional contribution to the total merger rate of each individual binary
formation channel is still unknown. The direct detection of an ElectroMagnetic
(EM) counterpart of a BBH merger would be optimal to identify its host galaxy.
The identification of candidate EM counterparts of mergers from AGN discs
have been claimed (\citealp{graham20,graham23}\citealp[, however, see also][]{ashton21}),
and several works have investigated what should be the features of such
counterparts \citep{palenzuela10,loeb16,bartos16,mckernan19,petrov22}.
However, the current observational evidence based on EM counterparts is
still not sufficient to constrain what fraction of the detected BBH mergers
come from a specific channel.

Besides the search for EM counterparts, another method to investigate the
contribution of a formation channel to the total detected merger rate is to
infer how the distributions of the parameters of the merging binary should be
for that specific formation path, and then compare these predictions to the
data obtained by the LIGO and Virgo interferometers. This approach has been
utilised in several previous works focused on the eccentricity of the binary
\citep{romeroshaw21,romeroshaw22,samsing22}, the components' spin orientation
\citep{vajpeyi22}, the components' mass distribution
\citep{gayathri21, gayathri23,belczynski22,stevenson22}, its redshift dependence
\citep{karathanasis22}, and its relation with the distribution of the magnitude and
the orientation of the spins \citep{mckernan20,qin22,wang22,zevin22}.
These works agree in saying that BBHs that merge in a dynamical environment tend to
have higher masses involved, and more isotropically orientated spins. However,
there is still no general agreement on the relative contributions to the total
merger rate of all the possible formation channels.

Finally, a promising possibility to directly infer the fraction of the observed
GW events that happened in a specific host environment is through the investigation
of the spatial correlation between GW sky maps and the positions of such potential
hosts. The statistical power of this approach has been investigated using simulated
data, finding that it is possible to put constraints on the fraction of observed GW
events that happened in an AGN, ($f_{\rm AGN}$), especially when rare (i.e. very
luminous) potential sources are taken into account \citep{bartos17,corley19,veronesi22}.
These previous works used as main inputs the size of the 90 per cent Credibility Level
localisation volume (further referred to as V90) of each GW observation and the
number of AGN within it.

In this work we put for the first time upper limits on $f_{\rm AGN}$, based
on the observed GW-AGN spatial correlation in the case of high-luminosity AGN.
These upper limits are obtained through the application of a statistical
method that uses for the first time as input the exact position of every AGN.
The likelihood function $\mathcal{L}\left(f_{\rm AGN}\right)$ described in
Section \ref{sec:like} takes also into account the incompleteness that
characterizes the catalogue of potential hosts. We implement a likelihood
maximization algorithm and check its performance on 3D Gaussian probability
distributions as emulators of GW sky maps, and a mock catalogue of AGN. We then
apply this method to check the spatial correlation between the objects of three
all-sky catalogues of observed AGN and the 30 BBH mergers, with a 90\%
Credible Interval (CI) on the redshift posterior distribution fully contained
within $z=0.3$. Every AGN catalogue is characterized by a different lower
cut in bolometric luminosity.

This paper is organized as follows: in Section \ref{sec:dataset} we describe the
properties of the observed all-sky AGN catalogues and of the detected GW events
our statistical method is applied on. In the same section, we report how we generate
the AGN mock catalogue and the Gaussian probability distributions necessary to
test the likelihood performance. In Section \ref{sec:method} we describe in detail
the analytical form of the likelihood function, how we test it on the mock AGN
catalogue, and how we apply it to real data. In Section \ref{sec:res} we present
the results of this application and the constraints on $f_{\rm AGN}$ it produces.
Finally, in Section \ref{sec:concl} we draw conclusions from these results and discuss
how they can be improved and generalised in the near future.

We adopt the cosmological parameters of the Cosmic Microwave Background
observations by Planck \citep{planck15}: $H_0=(67.8\pm0.9)\ {\rm km}\ {\rm s}^{-1}
{\rm Mpc}^{-1}$, $\Omega_{\rm m}=0.308\pm0.012$, $n_{\rm s}=0.968\pm0.006$.


\vspace{-0.5em}
\section{Datasets}
\label{sec:dataset}

In this section we first describe the selection criteria that we adopt to
build the three all-sky catalogues of observed AGN, and we present the 30
detected GW events used when applying our statistical method to real data.
We then describe the creation of the AGN mock catalogue and of the 3D Gaussian
probability distributions used to validate our statistical method.


\subsection{AGN catalogues}
\label{sec:agn}
In order to construct our AGN catalogues, we start from the unWISE
catalogue \citep{schlafly19}, which is based on the images from the WISE survey
\citep{wright10}, and cross-match it with version 7.7b of the Milliquas
catalogue \citep{flesch21}. This Milliquas catalogue puts together
all quasars from publications until October 2022, and contains a total of
2,970,254 objects. The cross-match is performed to associate a spectroscopic
redshift measurement to as many unWISE objects as possible. We then select
the objects with redshift estimates of $z\leq0.3$.
The reason in favour of restricting our analysis to $z\leq0.3$ is that
the constraining power of our approach scales linearly with the completeness
of the AGN catalogue that is used, and this redshift cut allows us
to have an AGN completeness $\gtrsim 0.5$.

\begin{figure*}
    \centering
    \includegraphics[trim= 45 80 20 110,clip,width=2\columnwidth]{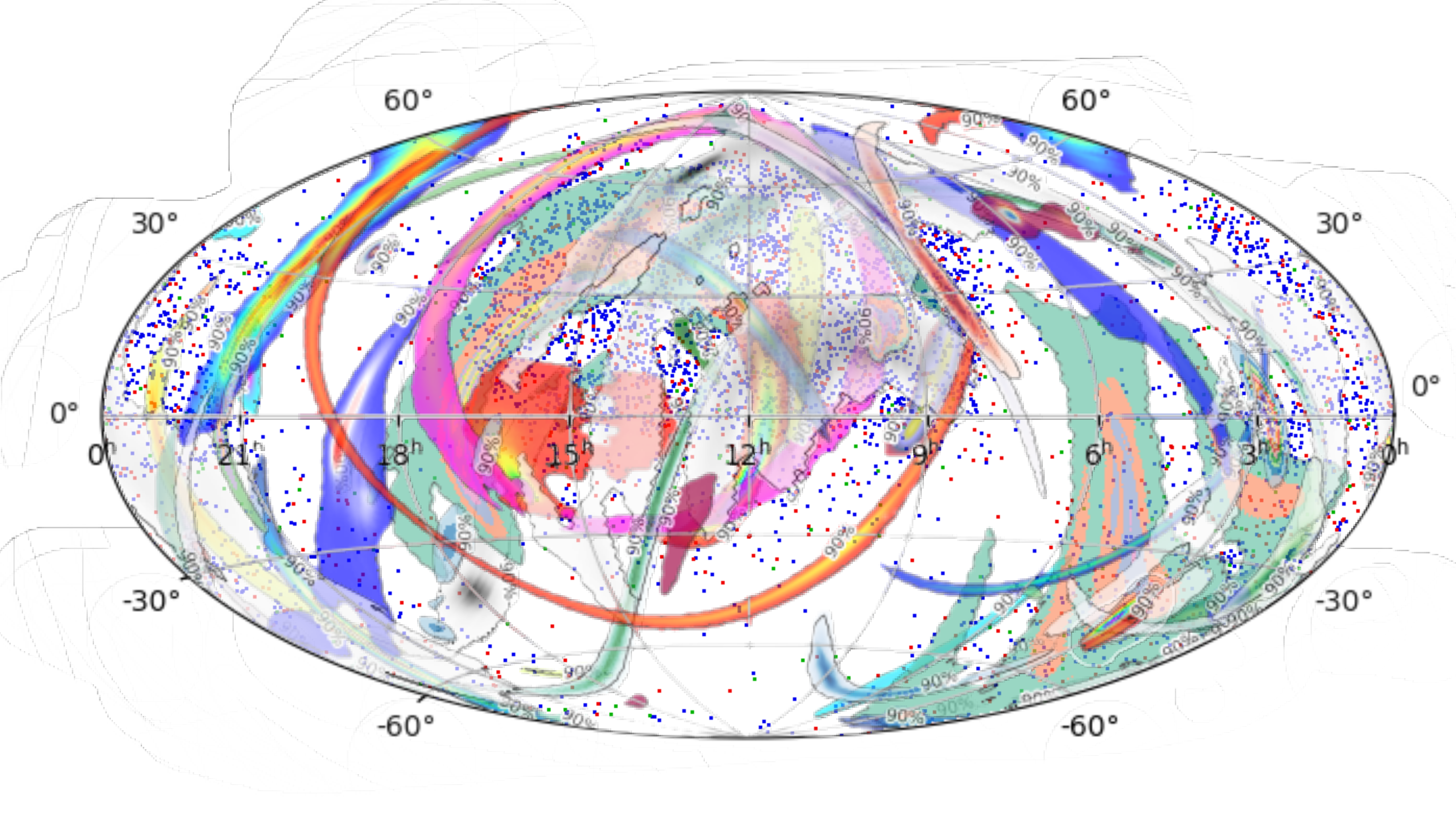}
    \caption{Positions of the AGN in CAT450 (blue dots), CAT455 (red dots),
    and CAT460 (green dots) described in Section \ref{sec:agn}, and 90 per
    cent CL localisation surfaces of the 30 detected BBH mergers listed in 
    \ref{tab:closegw}. These have a 90 per cent CI of the redshift
    posterior fully contained within $z=0.3$ (coloured regions). Regions with
    different colours correspond to different events. The sky map is visualized
    in equatorial coordinates.}
    \label{fig:skymaps}
\end{figure*}

\begin{figure}
    \centering
    \includegraphics[trim= 10 0 30 35,clip,width=1.\columnwidth]{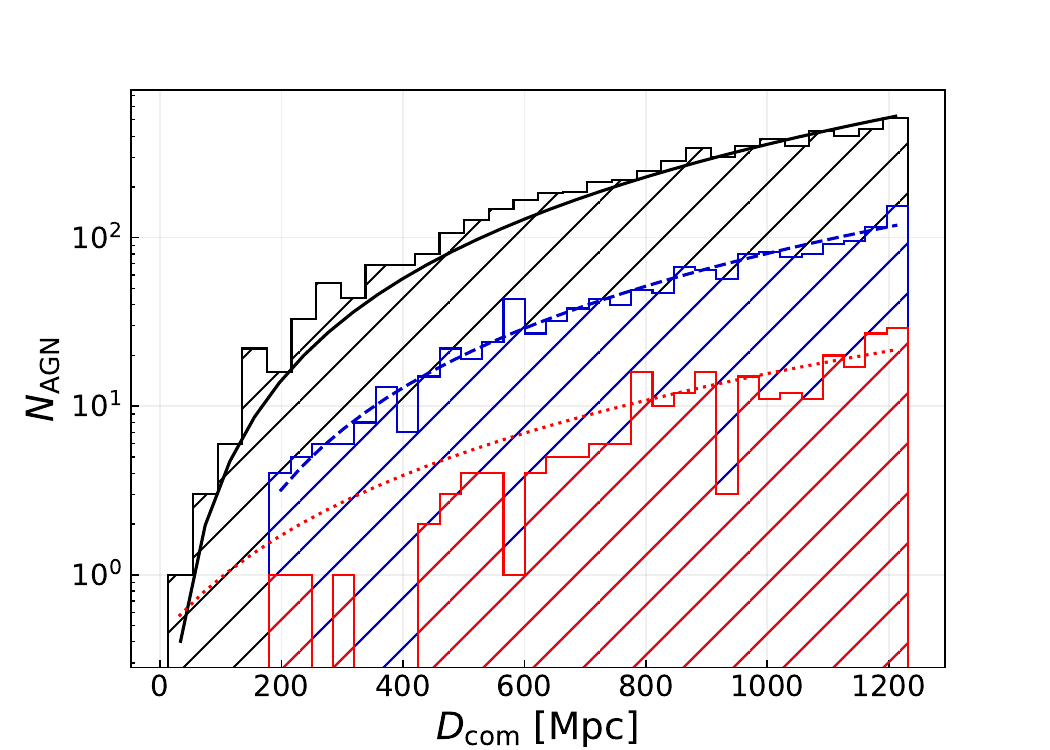}
    \caption{Number of AGN $N_{\rm AGN}$ in the catalogues presented in Section
    \ref{sec:agn} as a function of comoving distance $D_{\rm com}$. The black,
    blue, and red histograms refer to CAT450, CAT455, and CAT460, respectively. The black
    solid line, the blue dashed one, and the red dotted one show the best fit
    functions we obtain when fitting the number of objects per bin using the following
    form: $N_{\rm AGN}\propto D_{\rm com}^2$. These fits show no evidence of a
    significant redshift-dependent incompleteness of the catalogues. The apparent
    dearth of objects with $D_{\rm com}\leq400$ Mpc in CAT460 can be explained in terms
    of a random statistical fluctuation with respect to the expectation value.}
    \label{fig:comphist}
\end{figure}

We then use the flux in the W1 band of the WISE survey to calculate the
bolometric luminosity of every object and select only the ones brighter
than the luminosity threshold that characterizes each of the three catalogues
we create. These thresholds are $10^{45}\,{\rm erg\,s}^{-1}$,
$10^{45.5}\,{\rm erg\,s}^{-1}$, and $10^{46}\,{\rm erg\,s}^{-1}$.
Finally, we perform a color selection. We select objects with
${\rm mag(W1)-mag(W2)}\geq 0.8$, where ${\rm mag(W1)}$ is the magnitude in
the W1 band and ${\rm mag(W2)}$ is the magnitude in the W2 band. This is
done to select objects based on their features related to thermal emission
from hot dust, filtering out any contribution from the host galaxy to the AGN
luminosity \citep{assef13}. Such a selection is has been proven to lead
to a catalogue characterized by a reliability not smaller than 95 per cent
\citep{stern12}. The resulting contamination fraction lower than 5 per cent
is not expected to bias our results in a significant way.
In the lowest luminosity threshold catalogue, this colour cut removes
$\approx 62$ per cent of all AGN, while this percentage drops to $\approx 5$
per cent and $\approx 2$ per cent for the $10^{45.5}\,{\rm erg\,s}^{-1}$ and
$10^{46}\,{\rm erg\,s}^{-1}$ threshold catalogues, respectively. 
We are left with three catalogues containing 5,791, 1,412, and 242 AGN
for the bolometric luminosity thresholds of $10^{45}\,{\rm erg\,s}^{-1}$,
$10^{45.5}\,{\rm erg\,s}^{-1}$, and $10^{46}\,{\rm erg\,s}^{-1}$, respectively.
These three catalogues will be further referred to as CAT450, CAT455, and CAT460.
The two catalogues characterized by the two highest luminosity thresholds
are both subsamples of CAT450.
Even if the AGN in the catalogues are not uniformly distributed in the
sky (see Figure \ref{fig:skymaps}), they show no significant redshift-dependent
incompleteness. This can be established by checking that the number
of AGN ($N_{\rm AGN}$) in a specific bin of comoving distance ($D_{\rm com}$)
is proportional to $D_{\rm com}^2$ up to the maximum redshift of the catalogues:
$z=0.3$ (see Figure \ref{fig:comphist}).
A simple three-regions partition of the catalogues is used to identify areas with
similar 2D sky-projected number density of AGN. For CAT455 we have that:
\begin{itemize}
    \item 809 objects are within the footprint of the
    seventeenth data release of the Sloan Digital Sky Survey (SDSS)
    \citep{york00,blanton17,abdurro22} (which corresponds approximately to 35.28 per cent
    of the sky). This is the most crowded region of the three, with a 2D number
    density of $\approx 0.0556$ objects per square degree;
    \item 41 objects are characterized by a galactic latitude $b$ with an absolute
    value smaller than $10^\circ$ (approximately 17.36 per cent of the sky).
    In this region the Galactic plane of the Milky Way prevents observations
    from detecting most of the extra-galactic content, and is therefore the least
    crowded region of our catalogue, with 2D number density of $\approx 0.0057$
    objects per square degree;
    \item The remaining 562 objects populate the remaining 47.36 per cent of the sky.
    The average 2D number density in this region is $\approx 0.0288$ objects per square
    degree.
\end{itemize}
Because the AGN we consider and their host galaxies are relatively bright, many of
them fall within the flux limit of the SDSS spectrocopic galaxy sample \citep{strauss02},
which has a completeness close to 100 per cent. In addition, the SDSS spectroscopic
target selection \citep{richards02} is tuned to target AGN or quasars below this flux
limit. For this reason, the completeness of our catalogues in the SDSS footprint can
be assumed to be close to 100 per cent. We calculate the incompleteness of the other
two regions from the ratio of the projected 2D densities. Small deviations from unity
for the completeness in the SDSS footprint are not expected to significantly change our
final results.
The same partition of the sky has been used to estimate the completeness of CAT450
and CAT460. The estimated completenesses, weighted over the area occupied by
each region, are $\approx48$ per cent, $\approx61$ per cent, and $\approx87$ per cent 
for CAT450, CAT455, and CAT460, respectively.

We calculate the number densities of the AGN catalogues we create, correcting for their
completeness. We obtain a completeness-corrected number density of
$1.53\cdot10^{-6}{\rm Mpc}^{-3}$, $2.93\cdot10^{-7}{\rm Mpc}^{-3}$, and
$3.54\cdot10^{-8}{\rm Mpc}^{-3}$ for CAT450, CAT455, and CAT460, respectively. 
To illustrate the content of our catalogues, we show in Table \ref{tab:catalogue} as
an example the first ten entries of CAT450.

\begin{table*}
    \begin{adjustbox}{width=\linewidth,center}
    \begin{tabular}{c|c|c|c|c|c|c|c|c}
         Name & Citation for Name & unWISE ID &  R.A. &  Dec. & $z$ & Citation
         for $z$ & W1 mag & $L_{\rm W1}$\\
         & &  [deg] & [deg] & & & & & [${\rm erg\,s}^{-1}$]\\
         \hline
         UVQSJ000000.15-200427.7 & \citet{monroe16} & 0000m197o0005716 & $0.00065$ & $-20.07433$ &
         0.291 & \citet{monroe16} & $13.65$ & $2.72\cdot 10^{44}$\\
         SDSS J000005.49+310527.6 & \citet{ahumada20} & 0000p318o0001234 & $0.02290$ & $31.09102$ &
         0.286 & \citet{ahumada20} & $14.20$ & $1.58\cdot10^{44}$\\
         PHL 2525 & \citet{lamontagne00} & 0000m122o0001902 & $0.10172$ & $-12.76328$ &
         $0.200$ & \citet{lamontagne00} & $11.04$ & $1.29\cdot10^{45}$\\
         2MASX J00004028-0541012 & \citet{masci10} & 0000m061o0015237 & $0.16774$ & $-5.68361$ &
         $0.094$ & \citet{masci10} & $11.33$ & $1.90\cdot10^{44}$\\
         RXS J00009+1723 & \citet{wei99} & 0000p166o0024250 & $0.23319$ & $17.39413$ &
         $0.215$ & \citet{wei99} & $12.93$ & $2.64\cdot10^{44}$\\
         SDSS J000102.18-102326.9 & \citet{lyke20} & 0000m107o0014745 & $0.25911$ & $-10.39078$ &
         $0.294$ & \citet{lyke20} & $14.75$ & $1.01\cdot10^{44}$\\
         RX J00013+0728 & \citet{tesch00} & 0000p075o0010333 & $0.32534$ & $7.47432$ &
         $0.270$ & \citet{tesch00} & $14.06$ & $1.57\cdot10^{44}$\\
         PGC 929358 & \citet{paturel03} & 0000m137o0004668 & $0.33219$ & $-14.07310$ &
         $0.087$ & \citet{mauch07} & $11.65$ & $1.21\cdot10^{44}$\\
         PGC 1698547 & \citet{paturel03} & 0000p242o0009501 & $0.38474$ & $24.04179$ &
         $0.104$ & \citet{ahumada20} & $11.72$ & $1.65\cdot10^{44}$\\
         RX J00015+0529 & \citet{tesch00} & 0000p060o0003070 & $0.38896$ & $5.48926$ &
         $0.250$ & \citet{ahumada20} & $12.67$ & $4.71\cdot10^{44}$
    \end{tabular}
    \end{adjustbox}
        \caption{First ten objects from our publicly available catalogue of AGN
        with a bolometric luminosity higher than $10^{45}\,{\rm erg\,s}^{-1}$, in ascending
        order of Right Ascension. For every object we indicate the original ID from the
        literature, the paper that first presented it, its unWISE ID, Right Ascension,
        Declination, redshift, the paper that first presented that redshift estimate, the
        magnitude in the W1 band, and the luminosity in the same band, $L_{\rm W1}$. We
        calculate the bolometric luminosity multiplying $L_{\rm W1}$ by a bolometric correction
        factor, approximated to 10 for this band and in the luminosity range we consider
        \citep{hopkins07}. Out of the 5,791 objects in the catalogue, a total of 3,561 have
        a redshift measurement obtained from SDSS. In particular, 1,582 of these measurements
        are taken from \citet{lyke20}, 1,025 from \citet{ahumada20}, and 954 from
        \citet{liu19}. The full catalogue will be made available on the journal website
        and at \url{https://github.com/niccoloveronesi/AGNallskycat_Veronesi23.git}.}
    \label{tab:catalogue}
\end{table*}


\subsection{Detected Gravitational Wave events}
\label{sec:detgws}

When applying our statistical method to real data, we exploit the localisation
volumes of 30 BBH mergers. These were detected during the first three observing
runs of the LIGO and Virgo intereferometers. We select those with the
90 per cent CI of the redshift posterior distribution within $z=0.3$ and
false alarm rate below 1 per year. Our selected events are among the ones
used in \citet{ligo21pop} to infer the parameters of the sMBH astrophysical
population.
These sky maps have been downloaded from the Gravitational Wave Open Science Center
\citep{abbott21a}.

\begin{table*}
    \centering
    \begin{tabular}{c|c|c|c|c|c|c|c|c|c}
         \hline
         Event ID &  $m_1$ &  $m_2$ & $\chi_{\rm eff}$ & $z$ & SNR & ${\rm V90}$ &
         $N_{{\rm V90,CAT450}}$ & $N_{{\rm V90,CAT455}}$ & $N_{{\rm V90,CAT460}}$\\
         &  ${\rm M}_\odot$ & ${\rm M}_\odot$ & & & & $\left[{\rm Mpc}^3\right]$ & & &\\
         \hline
         GW150914\_095045 & $34.6_{-2.6}^{+4.4}$ & $30.0_{-4.6}^{+2.9}$ & $-0.04_{-0.14}^{+0.12}$
         & $0.10_{-0.03}^{+0.03}$ & $26.0_{-0.2}^{+0.1}$ & $3.39\cdot10^6$ & 3 & 0 & 0\\
         GW151226\_033853 & $14.2_{-3.6}^{+11.1}$ & $7.5_{-2.8}^{+2.4}$ & $0.20_{-0.08}^{+0.23}$
         & $0.10_{-0.04}^{+0.03}$ & $12.7_{-0.4}^{+0.3}$ & $1.32\cdot10^7$ & 10 & 1 & 0\\
         GW170104\_101158 & $28.7_{-4.2}^{+6.6}$ & $20.8_{-4.7}^{+4.1}$ & $-0.04_{-0.19}^{+0.15}$
         & $0.22_{-0.09}^{+0.07}$ & $13.8_{-0.3}^{+0.2}$ & $1.42\cdot10^8$ & 196 & 30 & 6\\
         GW170608\_020116 & $10.6_{-1.4}^{+4.0}$ & $7.8_{-1.9}^{+1.2}$ & $0.05_{-0.05}^{+0.13}$
         & $0.07_{-0.03}^{+0.03}$ & $15.3_{-0.3}^{+0.2}$ & $2.98\cdot10^6$ & 3 & 0 & 0\\
         GW170809\_082821 & $34.1_{-5.3}^{+8.0}$ & $24.2_{-5.3}^{+4.8}$ & $0.07_{-0.17}^{+0.17}$
         & $0.21_{-0.07}^{+0.05}$ & $12.8_{-0.3}^{+0.2}$ & $4.21\cdot10^7$ & 35 & 6 & 1\\
         GW170814\_103043 & $30.9_{-3.3}^{+5.4}$ & $24.9_{-4.0}^{+3.0}$ & $0.08_{-0.12}^{+0.13}$
         & $0.13_{-0.05}^{+0.03}$ & $17.7_{-0.3}^{+0.2}$ & $2.96\cdot10^6$ & 2 & 0 & 0\\
         GW170818\_022509 & $34.8_{-4.2}^{+6.5}$ & $27.6_{-5.1}^{+4.1}$ & $-0.06_{-0.22}^{+0.19}$
         & $0.21_{-0.07}^{+0.07}$ & $12.0_{-0.4}^{+0.3}$ & $6.04\cdot10^6$ & 3 & 1 & 0\\
         GW190412\_053044 & $27.7_{-6.0}^{+6.0}$ & $9.0_{-1.4}^{+2.0}$ & $0.21_{-0.13}^{+0.12}$
         & $0.15_{-0.04}^{+0.04}$ & $19.8_{-0.3}^{+0.2}$ & $9.16\cdot10^6$ & 20 & 3 & 0\\
         GW190425\_081805 & $2.1_{-0.4}^{+0.5}$ & $1.3_{-0.2}^{+0.3}$ & $0.07_{-0.05}^{+0.07}$
         & $0.03_{-0.01}^{+0.02}$ & $12.4_{-0.4}^{+0.4}$ & $7.78\cdot10^6$ & 9 & 1 & 0\\
         GW190630\_185205 & $35.1_{-5.5}^{+6.5}$ & $24.0_{-5.2}^{+5.5}$ & $0.10_{-0.13}^{+0.14}$
         & $0.18_{-0.07}^{+0.09}$ & $16.4_{-0.3}^{+0.2}$ & $1.23\cdot10^8$ & 148 & 33 & 4\\
         GW190707\_093326 & $12.1_{-2.0}^{+2.6}$ & $7.9_{-1.3}^{+1.6}$ & $-0.04_{-0.09}^{+0.10}$
         & $0.17_{-0.08}^{+0.06}$ & $13.1_{-0.4}^{+0.2}$ & $9.20\cdot10^7$ & 17 & 3 & 1\\
         GW190708\_232457 & $19.8_{-4.3}^{+4.3}$ & $11.6_{-2.0}^{+3.1}$ & $0.05_{-0.10}^{+0.10}$
         & $0.19_{-0.07}^{+0.06}$ & $13.4_{-0.3}^{+0.2}$ & $1.02\cdot10^9$ & 1560 & 305 & 43\\
         GW190720\_000836 & $14.2_{-3.3}^{+5.6}$ & $7.5_{-1.8}^{+2.2}$ & $0.19_{-0.11}^{+0.14}$
         & $0.16_{-0.05}^{+0.11}$ & $10.9_{-0.8}^{+0.3}$ & $4.24\cdot10^7$ & 20 & 7 & 1\\
         GW190725\_174728 & $11.8_{-3.0}^{+10.1}$ & $6.3_{-2.5}^{+2.1}$ & $-0.04_{-0.16}^{+0.36}$
         & $0.20_{-0.08}^{+0.09}$ & $9.1_{-0.7}^{+0.4}$ & $3.81\cdot10^8$ & 106 & 44 & 11\\
         GW190728\_064510 & $12.5_{-2.3}^{+6.9}$ & $8.0_{-2.6}^{+1.7}$ & $0.13_{-0.07}^{+0.19}$
         & $0.18_{-0.07}^{+0.05}$ & $13.1_{-0.4}^{+0.3}$ & $3.88\cdot10^7$ & 17 & 4 & 0\\
         GW190814\_211039 & $23.3_{-1.4}^{+1.4}$ & $2.6_{-0.1}^{+0.1}$ & $0.00_{-0.07}^{+0.07}$
         & $0.05_{-0.01}^{+0.01}$ & $25.3_{-0.2}^{+0.1}$ & $3.55\cdot10^4$ & 0 & 0 & 0\\
         GW190917\_033853 & $9.7_{-3.9}^{+3.4}$ & $2.1_{-0.4}^{+1.1}$ & $-0.08_{-0.43}^{+0.21}$
         & $0.15_{-0.06}^{+0.05}$ & $8.3_{-0.8}^{+0.5}$ & $1.05\cdot10^8$ & 60 & 22 & 3\\
         GW190924\_021846 & $8.8_{-1.8}^{+4.3}$ & $5.1_{-1.5}^{+1.2}$ & $0.03_{-0.08}^{+0.20}$
         & $0.11_{-0.04}^{+0.04}$ & $12.0_{-0.4}^{+0.3}$ & $1.27\cdot10^7$ & 13 & 1 & 0\\
         GW190925\_232845 & $20.8_{-2.9}^{+6.5}$ & $15.5_{-3.6}^{+2.5}$ & $0.09_{-0.15}^{+0.16}$
         & $0.19_{-0.07}^{+0.08}$ & $9.7_{-0.6}^{+0.3}$ & $2.86\cdot10^8$ & 401 & 94 & 12\\
         GW190930\_133541 & $14.2_{-4.0}^{+8.0}$ & $6.9_{-2.1}^{+2.4}$ & $0.19_{-0.16}^{+0.22}$
         & $0.16_{-0.06}^{+0.06}$ & $9.7_{-0.5}^{+0.3}$ & $1.32\cdot10^8$ & 63 & 13 & 2\\
         GW191103\_022549 & $11.8_{-2.2}^{+6.2}$ & $7.9_{-2.4}^{+1.7}$ & $0.21_{-0.10}^{+0.16}$
         & $0.20_{-0.09}^{+0.09}$ & $8.9_{-0.5}^{+0.3}$ & $3.16\cdot10^8$ & 255 & 62 & 8\\
         GW191105\_143512 & $10.7_{-1.6}^{+3.7}$ & $7.7_{-1.9}^{+1.4}$ & $-0.02_{-0.09}^{+0.13}$
         & $0.23_{-0.09}^{+0.07}$ & $9.7_{-0.5}^{+0.3}$ & $1.53\cdot10^8$ & 164 & 36 & 3\\
         GW191129\_134029 & $10.7_{-2.1}^{+4.1}$ & $6.7_{-1.7}^{+1.5}$ & $0.06_{-0.08}^{+0.16}$
         & $0.16_{-0.06}^{+0.05}$ & $13.1_{-0.3}^{+0.2}$ & $5.92\cdot10^7$ & 101 & 20 & 2\\
         GW191204\_171526 & $11.9_{-1.8}^{+3.3}$ & $8.2_{-1.6}^{+1.4}$ & $0.16_{-0.05}^{+0.08}$
         & $0.13_{-0.05}^{+0.04}$ & $17.5_{-0.2}^{+0.2}$ & $1.24\cdot10^7$ & 12 & 1 & 0\\
         GW191216\_213338 & $12.1_{-2.3}^{+4.6}$ & $7.7_{-1.9}^{+1.6}$ & $0.11_{-0.06}^{+0.13}$
         & $0.07_{-0.03}^{+0.02}$ & $18.6_{-0.2}^{+0.2}$ & $3.66\cdot10^6$ & 2 & 2 & 0\\
         GW200115\_042309 & $5.9_{-2.5}^{+2.0}$ & $1.44_{-0.29}^{+0.85}$ & $-0.15_{-0.42}^{+0.24}$
         & $0.06_{-0.02}^{+0.09}$ & $11.3_{-0.5}^{+0.3}$ & $3.79\cdot10^6$ & 3 & 2 & 0\\
         GW200129\_065458 & $34.5_{-3.2}^{+9.9}$ & $28.9_{-9.3}^{+3.4}$ & $0.11_{-0.16}^{+0.11}$
         & $0.18_{-0.07}^{+0.05}$ & $26.8_{-0.2}^{+0.2}$ & $7.06\cdot10^6$ & 7 & 0 & 0\\
         GW200202\_153413 & $10.1_{-1.4}^{+3.5}$ & $7.3_{-1.7}^{+1.1}$ & $0.04_{-0.66}^{+0.13}$
         & $0.09_{-0.03}^{+0.03}$ & $10.8_{-0.4}^{+0.2}$ & $2.32\cdot10^6$ & 2 & 0 & 0\\
         GW200311\_115853 & $34.2_{-3.8}^{+6.4}$ & $27.7_{-5.9}^{+4.1}$ & $-0.02_{-0.20}^{+0.16}$
         & $0.23_{-0.07}^{+0.05}$ & $17.8_{-0.2}^{+0.2}$ & $5.94\cdot10^6$ & 7 & 2 & 0\\
         GW200316\_215756 & $13.1_{-2.9}^{+10.2}$ & $7.8_{-2.9}^{+1.9}$ & $0.13_{-0.10}^{+0.27}$
         & $0.22_{-0.08}^{+0.08}$ & $10.3_{-0.7}^{+0.4}$ & $9.22\cdot10^7$ & 12 & 5 & 0\\
         \hline
    \end{tabular}
    \caption{List of the 30 BBH mergers detected during the first three
    observing runs of the LIGO and Virgo intereferometers with a CI of the redshift
    posterior contained within $z=0.3$ and a false alarm rate below
    1 per year.} For every event, we report its ID, the mass of both the primary ($m_1$)
    and the secondary ($m_2$) component, the effective inspiral spin parameter $\chi_{\rm eff}$
    \citep{ajith11}, the redshift, the SNR, and the value of V90. The last three columns
    correspond to the number of AGN inside V90, belonging to our three catalogues. We report
    the median and the 90 per cent credible intervals for the masses, the effective spin
    parameter, the redshift, and the SNR.
    \label{tab:closegw}
\end{table*}
Table \ref{tab:closegw} lists these events. Among the parameters we
report for each event, three are intrinsic properties of the binary.
These are the masses of the two components of the binary, and the effective
inspiral spin parameter. The latter is a weighted average of the projections
of the two components' spins on the direction of the angular momentum of
the binary \citep[for a more detailed description of this parameter,
see][]{ajith11,ligo21,ligo21pop}. The other parameters reported for each detected
GW event in Table \ref{tab:closegw} are the redshift, the SNR, V90, and
the number of AGN from our all-sky observed catalogues that are inside V90.
The 90 per cent CL sky regions of the same BBH mergers that are listed in Table
\ref{tab:closegw} are displayed in Figure \ref{fig:skymaps}.


\subsection{AGN mock catalogue}
\label{sec:agnmock}

We test our statistical method explained below on an AGN mock catalogue
characterized by a non-uniform incompleteness. In order to create it, we
first have to construct a {\it complete} parent mock catalogue, where
we assume that all AGN are accounted for. These are uniformly distributed
in comoving volume between $z=0.0$ and $z=0.4$ with a number density of
$n_{\rm AGN}= 10^{-7}{\rm Mpc}^{-3}$.
The non-uniform {\it incomplete} catalogue is a sub-sample of this complete one.
Non-uniform incompleteness is a feature present also in the observed
AGN catalogues exploited in this paper (see section \ref{sec:agn}). The
incomplete mock catalogue is created by dividing the complete one in
three different regions, and sub-sampling each of them in a different
way as follows:
\begin{itemize}
    \item The first region has galactic coordinate $b$ bigger than $30^\circ$.
    This corresponds to 25 per cent of the sky. In this first region no sub-sampling
    has been performed, hence its completeness is 100 per cent;
    \item The second region has $b$ between $-30^\circ$ and $30^\circ$. This
    corresponds to 50 per cent of the sky. In this second region, we remove 30 per
    cent of the objects from the parent complete catalogue, hence the completeness
    in this region is 70 per cent.
    \item The third region has Galactic coordinate $b$ smaller than $-30^\circ$.
    This corresponds to the remaining 25 per cent of the sky. Here we removed
    the 70 per cent of the objects from the complete catalogue, so the completeness
    of this region is 30 per cent.
\end{itemize}
The incomplete mock catalogue has a total of 1,160 objects, and a weighted
average completeness of 67.5 per cent.
 

\subsection{Simulated Gravitational Wave sky maps}
\label{sec:simulgws}

The sky maps of our simulated GW events are described for simplicity as 3D
Gaussian probability distributions. These distributions are created such that the
{\it size} of their 90 per cent Credibility Level volume is the same as the size of
an actual V90 simulated with the same source parameters, assuming the O3 configuration
of the LIGO and Virgo detectors.
For these simulated events we assume a Black Hole mass distribution that follows
the {\scshape Power Law + Peak} model described in \citet{ligo21pop}.
For simplicity the spins of the components of the binaries are assumed to be aligned with
the binary angular momentum, with a magnitude uniformly distributed between
$0$ and $1$. This choice does not bias our analysis. This is because assuming
aligned spins leads to distributions of V90 consistent with the observed one
\citep{veronesi22}. The size of V90 is the only parameter of the simulated BBH
merger detections that enters the analysis presented in this paper, together with
the spatial position.
The inclination $\iota$ of the binaries is sampled from a uniform distribution
in $\arccos{\iota}$.
Once we have sampled the distributions of all the parameters of the merging
BBH (masses and spins of the components, position of the merger and
inclination of the binary), we model its GW signal with an IMRPhenomD waveform
type \citep{husa16,khan16}. We then simulate the detection of this signal
with a network composed of three interferometers: LIGO Hanford, LIGO Livingston,
and Virgo. The sensitivity curves we use for these three detectors are the ones
correspondent to the following IDs: {\scshape aLIGOMidLowSensitivityP1200087}
for the LIGO interferometers, and {\scshape AdVMidLowSensitivityP1200087}
for Virgo. The duty cycle indicates for what fraction of the total observing
time each of the detectors
is online. To all detectors, we assign the average value of the duty cycles that
characterized the third observing run of LIGO and Virgo: 0.78
\citep{abbott21,ligo21}. We keep a Signal to Noise Ratio (SNR) detection
threshold of 8 for the network, and require SNR$\geq4$ for at least two of the
three detectors. This cut leads to a realistic distribution of V90 \citep{veronesi22},
allowing us to circumvent the need to calculate the detection confidence level,
according to the LIGO-Virgo-KAGRA collaboration criteria. \footnote{These are based
on the False Alarm Rate or the probability of being of astronomical origin,
$p_{\rm astro}$ \citep{abbott21,ligo21}.} We finally measure V90 for every
simulated detection using the {\scshape Bayestar} algorithm \citep{singer16}.
The sensitivity curves used to create these simulated detections and the value
chosen for the duty cycles aim to reproduce the network that performed real
detections during the third observing run of the LIGO and Virgo interferometers (O3).
However, we apply the method these simulations are used to test also to GW events
detected before O3. This does not introduce any bias in the testing strategy
described in Section \ref{sec:test}, because there is no V90 from the first
and second observing runs which is smaller (bigger) than the smallest (biggest)
one from O3 (see Table \ref{tab:closegw}).

To each simulated detection we therefore associate a value of V90. We call
R90 the radius of a sphere of volume V90. The 3D spherically symmetric
Gaussian distributions we use as mock GW sky maps are combinations of
three 1D Gaussian distributions with equal standard deviation.
For every value of R90, we calculate the standard deviation each of the 1D
distributions must have in order for the 90 per cent credibility contour
of the 3D Gaussian distribution to be a spherical surface of radius R90.

Knowing the exact position of each GW event we simulate, we can then sample
the coordinates of the centre of the correspondent mock sky map from a Gaussian
distribution centered on it. The standard deviation of such Gaussian is
calculated from the value of R90 associated to the simulated BBH merger.

The sample of mock sky maps for the testing of our statistical method is therefore
represented by 3D Gaussian distributions characterized by the positions of their
centres and the radii of their 90 per cent credibility level regions (R90).
The test strategy described in detail later in Section \ref{sec:test} is independent
on the shape of the sky maps used during the cross match with mock AGN catalogues. For
this reason, the choice of using a 3D Gaussian distribution does not lead to any bias
in the obtained results concerning the test of the validity of the statistical method.


\section{Method}
\label{sec:method}


\subsection{Likelihood function}
\label{sec:like}

Our statistical framework compares two scenarios.
In the first scenario AGN are physically associated to BBH mergers, while in the second
one, AGN are background sources, i.e, their presence inside the the localisation
volume of a GW event is coincidental.

The general analytical form of the likelihood function used in this work is based
on the one described in \citet{braun08} and first used to draw
conclusions on the detectability of a GW-AGN connection by \citet{bartos17}.
This can be written as follows:
\begin{align}
    \mathcal{L}\left(f_{\rm AGN}\right)&=\prod_{i=1}^{N_{\rm GW}}\mathcal{L}_i\left(f_{\rm AGN}\right) \nonumber \\
    &=\prod_{i=1}^{N_{\rm GW}}\left[c\cdot0.9\cdot f_{\rm AGN}\cdot\mathcal{S}_i+
    \left(1-c\cdot0.90\cdot
    f_{\rm AGN}\right)\mathcal{B}_i\right]\ \ ,
    \label{eq:like}
\end{align}
where $\mathcal{L}_i$ is the single-event likelihood associated to the
$i$-th GW event, $f_{\rm AGN}$ is the fraction of GW events that originate
from an AGN, $N_{\rm GW}$ is the total number of GW events, $c$ is the average
\footnote{As we mention later on, we tested that working with an average
incompleteness over the whole catalogue gives indistinguishable (correct)
results with respect to accounting for a position-dependent incompleteness.}
completeness of the AGN catalogue, and $\mathcal{S}_i$ ($\mathcal{B}_i$)
is the signal (background) probability density function. If the value
of $\mathcal{S}_i$ is bigger than the value of $\mathcal{B}_i$,
$\mathcal{L}_i\left(f_{\rm AGN}\right)$ will peak at the maximum allowed value
of its parameter: $f_{\rm AGN}=1$, meaning that the $i$-th GW event is likely
physically associated to one of the AGN that are inside its localisation volume.
The opposite is true if if the value of $\mathcal{B}_i$ is bigger than the value of
$\mathcal{S}_i$. The product of all the single-event likelihoods is then what
determines the degree of GW-AGN association through the value of $f_{\rm AGN}$
corresponding to its maximum. The $0.9$ pre-factor in front of $f_{\rm AGN}$
is used to take into account  that the localisation volumes we use are associated
to a confidence level of 90 per cent. The introduction of the $c$
factor is a novelty with respect to previous similar works that used only complete
mock AGN catalogues \citep{bartos17, corley19, veronesi22}. If such a term was not present
when using incomplete catalogues, the likelihood function would on average peak
at a lower value of $f_{\rm AGN}$ with respect to the true one. This would happen
because, even if a physical association exists, it might not be detected if the AGN
host of a GW event is not present in the catalogue. The $c$ factor in Equation \ref{eq:like}
corrects for this potential bias.
Previous studies used as main input the size of each GW event's
V90 and the number of AGN within it ($N_{{\rm V90}}$). In this work, we
additionally exploit the information embedded in the exact position of every
AGN within the localisation volume: i.e., the value of the 3D GW localisation
probability density function at the AGN position.
We therefore write the signal probability density function for the $i$-th GW as:
\begin{equation}
    \mathcal{S}_i=\frac{\sum_{j=1}^{N_{{\rm V90}_i}}p_j}{n_{\rm AGN}
    {\rm V90}_i}\ \ ,
    \label{eq:Si}
\end{equation}
where $n_{\rm AGN}$ is the average number density of AGN in the catalogue, and
$p_j$ is the probability density associated to the position of the $j$-th AGN.
The denominator in Equation \ref{eq:Si} represents the expected number
of AGN from a catalogue of number density $n_{\rm AGN}$ that are contained
in a region of size ${\rm V90}_i$. Therefore, the signal probability density
function represents the total probability density associated to the positions
of all the AGN within ${\rm V90}_i$, normalized by their expected number.
The more objects there are within ${\rm V90}_i$ and/or the more clustered they
are towards the 
peak of the probability density distribution, the higher the value of
$\mathcal{S}_i$ is. This is in accord with the fact that $\mathcal{S}_i$ in
Equation \ref{eq:Si} describes how likely the scenario in which AGN are
physically associated to BBH mergers is.
On the other hand, the probability density function associated to the scenario
where AGN are background sources, accidentally present in GW localisation volumes,
can be expressed with a flat probability for an AGN to be found anywhere in V90:
\begin{equation}
    \mathcal{B}_i=\frac{0.9}{{\rm V90}_i}\ \ ,
    \label{eq:Bi}
\end{equation}
where the $0.9$ term at the numerator guarantees that $\mathcal{S}_i$ and
$\mathcal{B}_i$ are normalized to the same value.
From Equations \ref{eq:Si} and \ref{eq:Bi} it follows that the likelihood
function in Equation \ref{eq:like} is dimensionful with units of one over volume.
This means that for it to be turned into a probability density function, it
should be normalized dividing it by its integral over the whole [0,1] range
of $f_{\rm AGN}$. During the testing of the statistical method on mock data
and its application to real GW detections and AGN catalogues the non-normalized
version of the likelihood function is usually computed, unless specified otherwise.
In particular we normalize this function when extracting the posterior distribution
on $f_{\rm AGN}$.

In our statistical analysis the prior on $f_{\rm AGN}$ is assumed to be uniform
between 0 and 1.


\subsection{Test on mock data}
\label{sec:test}
To test the performance of the likelihood we use data coming from the cross-match
between the incomplete AGN mock catalogue described in Section \ref{sec:agnmock}
and the mock GW detections described in Section \ref{sec:simulgws}.

This test consists of a Monte Carlo simulation of 1,000 realizations. Every realization
is characterized by the same total number of simulated detected BBH mergers. This number of
detections is the same one used during the application to real data: $N_{\rm GW}=30$. 
At the start of each realization, we draw a value from the prior distribution of
$f_{\rm AGN}$. This represents the true value of this parameter for the specific realization,
and will be further referred to as $f_{\rm AGN,true}$. We then sample a binomial distribution
characterized by the parameters $n=N_{\rm GW}$ and $p=f_{\rm AGN,true}$ to obtain the
number of simulated detected GWs that come from an AGN of the
complete mock catalogue presented in Section \ref{sec:agnmock} within $z=0.2$. The
remaining events of the $N_{\rm GW}$ simulated detections are the ones coming from a
position randomly sampled from a uniform distribution in the same redshift range.
The redshift cut on the potential sources of both the signal and the background events is
performed to be sure that the entirety of V90 is within the volume of the mock AGN catalogue.
This is necessary to avoid any boundary-related underestimation of $\mathcal{S}_i$ during
the cross-match of these localisation volumes with the incomplete AGN mock catalogue.

We cross-match the 3D Gaussian distributions representing the sky maps of the 30 GW events
with the incomplete AGN mock catalogue and calculate the value of the likelihood as a function
of $f_{\rm AGN}$ using Equations \ref{eq:like}, \ref{eq:Si}, and \ref{eq:Bi}. We then compute
the normalized posterior distribution on $f_{\rm AGN}$: $\mathcal{P}\left(f_{\rm AGN}\right)$.
Finally, we calculate the Credibility Level (CL) of $f_{\rm AGN, true}$ and the corresponding
Credibility Interval (CI).
The CI is defined as the range of $f_{\rm AGN}$ that is associated to values of the
posterior equal or greater than $\mathcal{P}\left(f_{\rm AGN, true}\right)$. We say for example
that $f_{\rm AGN, true}$ has a CL of 90 per cent if the integral of
$\mathcal{P}\left(f_{\rm AGN}\right)$ evaluated over the corresponding CI is $0.9$.

\begin{figure}
    \centering
    \includegraphics[trim= 10 0 30 55,clip,width=0.49\textwidth]{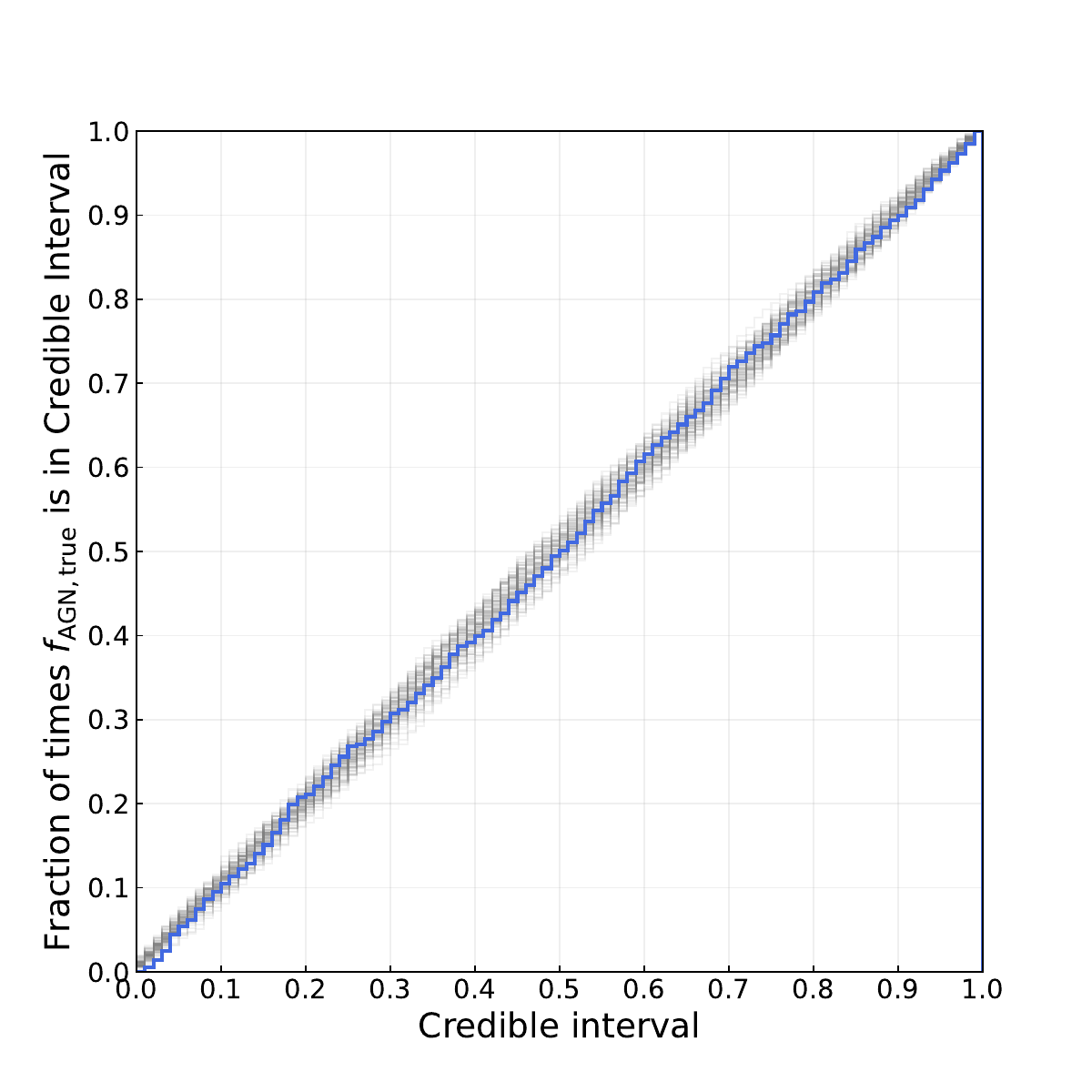}
    \caption{Fraction of times $f_{\rm AGN, true}$ lies within a certain Credible Interval
    as a function of the credibility level of such an interval. The blue line shows the
    result obtained by testing the likelihood function described in Section \ref{sec:like}
    on mock data. The gray lines show the cumulative distributions of 100 samples of
    a uniform distribution in the [0,1] range.}
    \label{fig:pp}
\end{figure}

The blue line in the Probabilty-Probability plot presented in Figure \ref{fig:pp}
shows the cumulative distribution of the 1,000 values of CLs associated to
$f_{\rm AGN,true}$ from all the realizations. The grey lines show the cumulative
distribution of 100 uniform samples between 0 and 1.
Since the distribution of the CLs associated to $f_{\rm AGN,true}$ is 
statistically indistinguishable from a uniform one, we can conclude that our
statistical method is able to produce trustworthy results when tested on mock data.
Therefore, maximizing the likelihood described in Equations \ref{eq:like}, \ref{eq:Si} and
\ref{eq:Bi} leads to an accurate estimate of $f_{\rm AGN}$.

Finally, we test that our results do not change if we use in Equation \ref{eq:like}
the actual value of the catalogue completeness ($c$) in each localisation volume.
More specifically, this individual completeness is calculated as a weighted average
of the completeness of the AGN catalogue in the 3D region occupied by each V90.
Our test yields indistinguishable results, therefore, for simplicity, we only
present the ones computed using the average catalogue completeness.


\subsection{Application to real data}
\label{sec:appl}
Once we have tested the accuracy of the statistical method, we apply it to real
data. We cross-match the skymaps of the 30 detected BBH mergers presented in
\ref{sec:detgws}, and listed in Table \ref{tab:closegw} with the all-sky AGN
catalogues described in Section \ref{sec:agn}.
We then calculate $\mathcal{L}\left(f_{\rm AGN}\right)$ using Equations
\ref{eq:like}, \ref{eq:Si}, and \ref{eq:Bi}.

In the case of CAT455 and CAT460 the combination of the data coming from the
cross-match with the 30 GW events leads to a monotonically decreasing likelihood,
as a function of $f_{\rm AGN}$. We therefore decide to evaluate upper limits on
this parameter integrating the normalized likelihood between $f_{\rm AGN}=0$ and
$f_{\rm AGN}=1$. Since the prior is assumed to be uniform, through this integration
we obtain the cumulative posterior distribution on $f_{\rm AGN}$.

The same process has been followed also for CAT450, even if in this
case the likelihood turns out to be rather insensitive to $f_{\rm AGN}$.
Specifically, in this last case, the posterior is prior-dominated: data do not
allow us to put much tighter constraints on $f_{\rm AGN}$ than the ones imposed by
the flat prior only. This is caused by the high number of objects contained
in the AGN catalogue \citep{veronesi22}, combined with the non-negligible level of
incompleteness that characterizes the same catalogue. We therefore decide not to
repeat the analysis with an AGN catalogue characterized by a lower luminosity
threshold. Such a catalogue would likely also show redshift-dependent completeness,
which will have to be taken into account in future works aimed to explore the
relation between BBH mergers and lower-luminosities AGN. A meaningful exploitation of
AGN catalogues denser than the ones used in this work will be possible only
when we will have data from more and/or better localized BBH mergers.


\vspace{-0.7em}
\section{Results}
\label{sec:res}

\begin{figure}
    \centering
    \includegraphics[trim= 10 0 30 55,clip,width=0.49\textwidth]{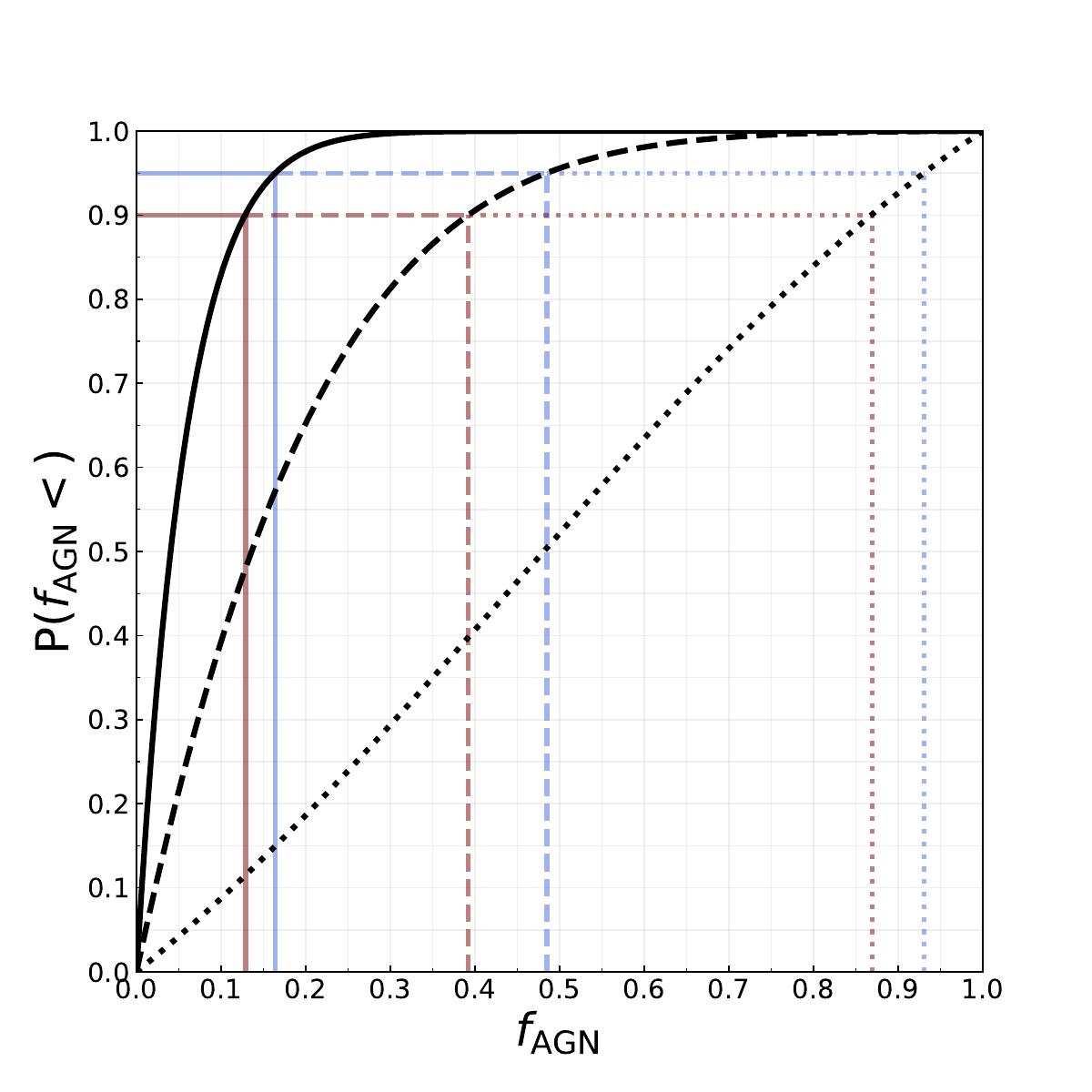}
    \caption{Black solid line: Cumulative posterior distribution
    for the fraction of detected GWs originated in an AGN ($f_{\rm AGN}$)
    with a bolometric luminosity higher than $10^{46}\,{\rm erg\,s}^{-1}$.
    Every value on the vertical axis corresponds to the probability associated
    to the true value of $f_{\rm AGN}$ being smaller than the correspondent
    value on the horizontal axis. The dashed (dotted) line shows the posterior
    distribution obtained using a luminosity threshold of $10^{45.5}\,{\rm erg\,s}^{-1}$
    ($10^{45}\,{\rm erg\,s}^{-1}$).
    The maroon lines indicate that the upper limit of the 90 per cent credibility
    interval corresponds to $f_{\rm AGN}=0.13$ for the $10^{46}\,{\rm erg\,s}^{-1}$ luminosity
    cut, to $f_{\rm AGN}=0.40$ for the $10^{45.5}\,{\rm erg\,s}^{-1}$ luminosity
    cut, and to $f_{\rm AGN}=0.87$ for the $10^{45}\,{\rm erg\,s}^{-1}$ luminosity
    cut. The blue lines indicate that the upper limit of the 95 per cent credibility
    interval corresponds to $f_{\rm AGN}=0.17$ for the $10^{46}\,{\rm erg\,s}^{-1}$ luminosity
    cut, to $f_{\rm AGN}=0.49$ for the $10^{45.5}\,{\rm erg\,s}^{-1}$ luminosity
    cut, and to $f_{\rm AGN}=0.94$ for the $10^{45}\,{\rm erg\,s}^{-1}$ luminosity
    cut.} 
    \label{fig:post}
\end{figure}

The cumulative posterior distributions over $f_{\rm AGN}$ we obtain through
the application of our statistical method to observed data are shown in Figure
\ref{fig:post}.
The black solid line shows the posterior distribution in the case of the
cross-math of the observed GW events with CAT460, while the dashed (dotted)
line shows it in the case of a CAT455 (CAT450).
On the vertical axis there is the probability for the true value of $f_{\rm AGN}$
being smaller than the correspondent value on the horizontal axis. As an example,
the solid blue line shows that the upper limit of the 95 per cent credibility
interval is $f_{\rm AGN}=0.17$ in the case of the cross-match with CAT460.
\begin{figure*}
    \centering
    \includegraphics[trim= 20 270 10 240,clip,width=2\columnwidth]{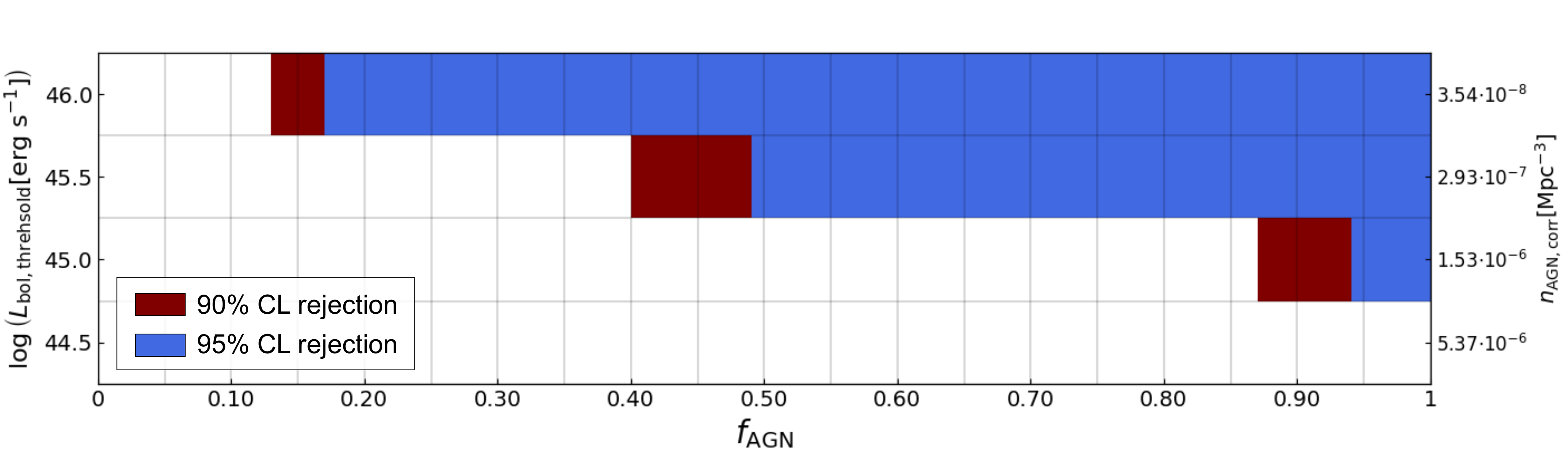}
    \caption{Rejected regions at 90 and 95 per cent credibility level of the two-dimensional
    parameter space $\{L_{\rm bol},f_{\rm AGN}\}$ investigated in this work.
    The bolometric luminosity threshold for the observed AGN is indicated on the vertical axis
    on the left-hand side, while the fraction of detected BBH mergers that come from
    AGN brighter than those thresholds is on the horizontal axis.
    The maroon (blue) regions are the ones that the analysis presented in this work
    rejects with a 90 (95) per cent credibility. The right vertical axis shows the
    number density obtained from the \citet{hopkins07} luminosity function, normalized to
    match the completeness-corrected number density of our catalogue.}
    \label{fig:parspace}
\end{figure*}
Figure \ref{fig:parspace} shows a region of the two-dimensional parameter space that has
been investigated in this work. On the vertical axis one can read the thresholds in
bolometric luminosities of AGN on the left-hand side, and the correspondent values
of number densities on the right-hand side. The three number densities correspondent
to the three luminosity thresholds we use to create CAT450, CAT455, and CAT460 have
been calculated taking into account their estimated completeness. For each of these
completeness-corrected number densities we calculate their ratio with respect to the
number density obtained integrating in the same luminosity range the best-fit AGN
luminosity function at $z=0.1$ presented in \citet{hopkins07}. The mean of this ratios,
together with the number density estimated from \citet{hopkins07} for a bolometric
luminosity threshold of $10^{44.5}\,{\rm erg\,s}^{-1}$, has been used to calculate the
completeness-corrected number density for such a luminosity cut.
All the possible values of $f_{\rm AGN}$ are on the horizontal axis. The maroon (blue) region
is the part of the parameter space that we reject with a 90 (95) per cent credibility level.

In \citet{ligo21pop} the total BBH merger rate per comoving volume has been
parametrized as a power law as a function of redshift: $\mathcal{R}(z)\propto(1+z)^\kappa$.
The value of the spectral index has been estimated to be $\kappa=2.7_{-1.9}^{+1.8}$,
and the best measurement of the merger rate $\mathcal{R}$ occurs at $z\approx0.2$:
$\mathcal{R}(z=0.2)\leq41\  {\rm Gpc}^{-3}{\rm yr}^{-1}$ at 90 per cent credibility.
Combining this result with the upper limit of $f_{\rm AGN}\leq0.49$ ($f_{\rm AGN}\leq0.17$)
obtained in this work, we find that the 95 per cent credibility upper limit on the rate
of BBHs merging in AGN brighter than $10^{45.5}\,{\rm erg\,s}^{-1}$ ($10^{46}\,{\rm erg\,s}^{-1}$) is
$\mathcal{R}_{\rm AGN}(z=0.2)\simeq 20\ {\rm Gpc}^{-3}{\rm yr}^{-1}$
($\mathcal{R}_{\rm AGN}(z=0.2)\simeq 7\ {\rm Gpc}^{-3}{\rm yr}^{-1}$) at $z\approx0.2$.
It is important to remember that these results have been obtained assuming
100 per cent completeness in the SDSS footprint in our catalogues of luminous, redshift
selected AGN. However, small variations over this assumption are not expected
to produce qualitatively different results with respect to the ones presented
in this section, since they scale linearly with the AGN catalogue completeness
(see Equation \ref{eq:like}).


\vspace{-0.5em}
\section{Discussion and conclusion}
\label{sec:concl}

We present a likelihood-based method to constrain the fractional contribution
of the AGN channel to the observed merger rate of BBHs. In particular we
compare the scenario in which AGN are physically associated to BBH mergers
to the one in which the presence of AGN in localisation volumes of GW events
is only due by chance.
We use as input data the size of each GW localisation volume and the exact
position of all the AGN that are in it. We calculate the posterior distribution
of the fraction of the detected GW events that come from an AGN, $f_{\rm AGN}$.
We then put observational constraints on this parameter by determining the
upper limits associated to the 90 and 95 per cent CIs of the
posterior distribution.

We first validate this method on a mock AGN catalogue characterized by a
non-uniform completeness (see Figure \ref{fig:pp}).

We then apply the same statistical analysis to observed data. We use the
sky maps of the 30 BBH mergers detected by the LIGO and Virgo interferometers
characterized by a 90 per cent C.I. of the redshift distribution entirely
contained within $z=0.3$. We cross-match these sky maps with three all-sky
catalogues of AGN we create starting from cross-matching the unWISE catalogue
\citep{schlafly19} with the Milliquas one \citep{flesch21}. We select only
the objects with a spectroscopic measurement of redshift correspondent to
$z\leq0.3$ and with a bolometric luminosity higher than $10^{45}\,{\rm erg\,s}^{-1}$,
$10^{45.5}\,{\rm erg\,s}^{-1}$, and $10^{46}\,{\rm erg\,s}^{-1}$. We calculate the
posterior cumulative distribution on $f_{\rm AGN}$ and conclude that in the
case of the two highest luminosity thresholds we can put upper limits on this
parameter that are tighter with respect to the ones one can obtain from the
sole assumption of a uniform prior between 0 and 1. In the case of the cross-match
with the AGN catalogue characterized by the highest (intermediate) luminosity
threshold we find that $f_{\rm AGN}=0.17$ ($f_{\rm AGN}=0.49$) is the upper
limit of the 95 per cent credibility interval.
Figure \ref{fig:post} shows the entire cumulative posterior distributions,
while Figure \ref{fig:parspace} shows more explicitly which parts of the
two-dimensional AGN luminosity-$f_{\rm AGN}$ parameter space are rejected with
a 90 and a 95 per cent credibility.
Previous works used only simulated GW data and mock AGN catalogues to draw conclusions
about the possibility of exploring the spatial correlation between the two. Instead,
we present the first constraints on $f_{\rm AGN}$ based on observational data only.
Moreover, in the previous analyses the number of potential hosts within the V90
of every GW event was used as the main source of information, together
with the size of V90. As mentioned above, the likelihood function we present in this
work also takes into account for the first time the exact position of every AGN within
V90 and the overall completeness of the AGN catalogue.
The results obtained in this work are observational upper limits on the
correlation between the detected BBH mergers and the high luminosity,
and spectroscopically selected AGN that are in the catalogues described
in Section \ref{sec:agn}. They can be used in the future to inform theoretical models
of compact binary objects in AGN discs. Such results hint towards the conclusion
that physical conditions of the gas and the stars in the discs of high-luminosity
AGN are not sufficiently able to drive the formation and the merger of binaries of
sMBHs in order to be major contributors to the total merger rate. This conclusion
would be in agreement with the recent theoretical result obtained by \citet{grishin23},
where it is stated that migration traps in AGN discs are not expected to be present
in the case of a bolometric luminosity higher than $10^{45}\ {\rm erg\ s}^-1$ for an AGN
alpha viscosity parameter of $\alpha=0.01$. Their inability to create
migration traps would explain why AGN characterized by a luminosity higher than
such a threshold are not to be considered potential preferred hosts of BBH mergers.

One way for generalizing the results presented in this paper is
the creation of a more complete all-sky AGN catalogue. The introduction
of objects with only a photometric measurement of the redshift is a
possible method of doing that. This would increase the number density of
the catalogue, but will also increase the probability of considering objects
that have been erraneously identified as AGN. This confidence on the
classification of each object will have to be taken into account in the
expression of the likelihood function.

The results concerning the posterior distributions shown in Figure \ref{fig:post}
are relative to the fraction of BBH mergers that have happened in an AGN
with a bolometric luminosity higher than the three thresholds we have considered.
We perform this luminosity cuts in order to be sure to have a good level of
completeness in our observed AGN catalogues.
In order to draw general conclusion on the AGN formation channel for BBHs,
future works will investigate the correlation between GW events and AGN in a
broader range of luminosities. Such an investigation will have to take into
consideration the fact that low values of complenetess and its dependence on redshift
lower the statistical power of the method, increasing the uncertainty on the predictions.

The analysis described in this paper is restricted to BBH mergers whose host
environment is expected to be at $z\leq0.3$ with 90 per cent credibility.
This selection has been done because a higher level of completeness for catalogues
of observed AGN can be reached if we restrict our analysis to the local Universe.
Future works might explore the GW-AGN correlation on a wider redshift range.
The effectiveness of their results will be increased because of the possible
exploitation of more detected BBH mergers, but might also be dampened by low levels
of completeness of the considered AGN catalogues.

Dedicated tests performed by varying the different parameters in the Monte 
Carlo analysis described in Section \ref{sec:test} have proven that the prediction
power of the method presented in this work depends mainly on three elements: the
completeness of the AGN catalogue, the number of GW detections, and the size of
their localisation volumes. Observational limitations (e.g. the presence of the
Milky Way plane that does not allow the detection of light coming from objects
behind it) prevent us from having an AGN catalogue with a completeness level close
to unity. On the other hand, $79^{+89}_{-44}$ BBH mergers are expected to be
observed via GWs during the fourth observing run (O4) of the LIGO-Virgo-KAGRA
collaboration \citep{abbott20}, and at least the same amount of detections can
be predicted for the fifth observing run (O5). This would at least triple the amount
of detected events available for statistical analyses on the BBH population.
This increase of the number of detections, together with the improvement on the
localisation power expected for O4 and O5 with respect to previous observing
runs, will noticeably increase the prediction power of likelihood-based methods
like the one presented in this paper, that will be able to put more
stringent constraints on the fractional contribution of high-luminosity AGN to
the total BBH merger rate, and to make use of also denser catalogues of
potential hosts, such as the ones containing AGN with luminosities lower than
the ones considered in this work.


\vspace{-0.8em}
\section*{Acknowledgements}
The authors thank Ilya Mandel for the stimulating discussion
regarding the how to assess the validity of the method when tested
on simulated data, and the anonymous referee for their comments
that helped to improve the presentation of our results.
EMR acknowledges support from ERC Grant ``VEGA P.", number 101002511.
This research has made use of data or software obtained from
the Gravitational Wave Open Science Center (gwosc.org), a service
of LIGO Laboratory, the LIGO Scientific Collaboration, the Virgo
Collaboration, and KAGRA. LIGO Laboratory and Advanced LIGO are
funded by the United States National Science Foundation (NSF) as
well as the Science and Technology Facilities Council (STFC) of
the United Kingdom, the Max-Planck-Society (MPS), and the State
of Niedersachsen/Germany for support of the construction of Advanced
LIGO and construction and operation of the GEO600 detector.
Additional support for Advanced LIGO was provided by the Australian
Research Council. Virgo is funded, through the European Gravitational
Observatory (EGO), by the French Centre National de Recherche Scientifique
(CNRS), the Italian Istituto Nazionale di Fisica Nucleare (INFN)
and the Dutch Nikhef, with contributions by institutions from Belgium,
Germany, Greece, Hungary, Ireland, Japan, Monaco, Poland, Portugal,
Spain. KAGRA is supported by Ministry of Education, Culture, Sports,
Science and Technology (MEXT), Japan Society for the Promotion of
Science (JSPS) in Japan; National Research Foundation (NRF) and
Ministry of Science and ICT (MSIT) in Korea; Academia Sinica (AS) and
National Science and Technology Council (NSTC) in Taiwan. 
{\em Software}: 
\texttt{Numpy} \citep{harris20}; 
\texttt{Matplotlib} \citep{hunter07}; 
\texttt{SciPy} \citep{virtanen20};
\texttt{Astropy} \citep{astropy13,astropy18};
\texttt{BAYESTAR} \citep{singer16}.

\vspace{-0.8em}
\section*{Data Availabilty}
The data underlying this article are available in niccoloveronesi/AGNallskycat\_Veronesi23,
at \url{https://github.com/niccoloveronesi/AGNallskycat_Veronesi23.git}.

\vspace{-1.0em}
\bibliographystyle{mnras}
\bibliography{bibliography}

\begin{thebibliography}{}
\makeatletter
\relax
\def\mn@urlcharsother{\let\do\@makeother \do\$\do\&\do\#\do\^\do\_\do\%\do\~}
\def\mn@doi{\begingroup\mn@urlcharsother \@ifnextchar [ {\mn@doi@}
  {\mn@doi@[]}}
\def\mn@doi@[#1]#2{\def\@tempa{#1}\ifx\@tempa\@empty \href
  {http://dx.doi.org/#2} {doi:#2}\else \href {http://dx.doi.org/#2} {#1}\fi
  \endgroup}
\def\mn@eprint#1#2{\mn@eprint@#1:#2::\@nil}
\def\mn@eprint@arXiv#1{\href {http://arxiv.org/abs/#1} {{\tt arXiv:#1}}}
\def\mn@eprint@dblp#1{\href {http://dblp.uni-trier.de/rec/bibtex/#1.xml}
  {dblp:#1}}
\def\mn@eprint@#1:#2:#3:#4\@nil{\def\@tempa {#1}\def\@tempb {#2}\def\@tempc
  {#3}\ifx \@tempc \@empty \let \@tempc \@tempb \let \@tempb \@tempa \fi \ifx
  \@tempb \@empty \def\@tempb {arXiv}\fi \@ifundefined
  {mn@eprint@\@tempb}{\@tempb:\@tempc}{\expandafter \expandafter \csname
  mn@eprint@\@tempb\endcsname \expandafter{\@tempc}}}

\bibitem[\protect\citeauthoryear{{Abbott} et~al.,}{{Abbott}
  et~al.}{2020}]{abbott20}
{Abbott} B.~P.,  et~al., 2020, \mn@doi [Living Reviews in Relativity]
  {10.1007/s41114-020-00026-9}, \href
  {https://ui.adsabs.harvard.edu/abs/2020LRR....23....3A} {23, 3}

\bibitem[\protect\citeauthoryear{{Abbott} et~al.,}{{Abbott}
  et~al.}{2021a}]{abbott21}
{Abbott} R.,  et~al., 2021a, \mn@doi [Physical Review X]
  {10.1103/PhysRevX.11.021053}, \href
  {https://ui.adsabs.harvard.edu/abs/2021PhRvX..11b1053A} {11, 021053}

\bibitem[\protect\citeauthoryear{{Abbott} et~al.,}{{Abbott}
  et~al.}{2021b}]{abbott21a}
{Abbott} R.,  et~al., 2021b, \mn@doi [SoftwareX] {10.1016/j.softx.2021.100658},
  \href {https://ui.adsabs.harvard.edu/abs/2021SoftX..1300658A} {13, 100658}

\bibitem[\protect\citeauthoryear{{Abdurro'uf} et~al.,}{{Abdurro'uf}
  et~al.}{2022}]{abdurro22}
{Abdurro'uf} et~al., 2022, \mn@doi [\apjs] {10.3847/1538-4365/ac4414}, \href
  {https://ui.adsabs.harvard.edu/abs/2022ApJS..259...35A} {259, 35}

\bibitem[\protect\citeauthoryear{{Acernese} et~al.,}{{Acernese}
  et~al.}{2015}]{acernese15}
{Acernese} F.,  et~al., 2015, \mn@doi [Classical and Quantum Gravity]
  {10.1088/0264-9381/32/2/024001}, \href
  {https://ui.adsabs.harvard.edu/abs/2015CQGra..32b4001A} {32, 024001}

\bibitem[\protect\citeauthoryear{{Ahumada} et~al.,}{{Ahumada}
  et~al.}{2020}]{ahumada20}
{Ahumada} R.,  et~al., 2020, \mn@doi [\apjs] {10.3847/1538-4365/ab929e}, \href
  {https://ui.adsabs.harvard.edu/abs/2020ApJS..249....3A} {249, 3}

\bibitem[\protect\citeauthoryear{{Ajith} et~al.,}{{Ajith}
  et~al.}{2011}]{ajith11}
{Ajith} P.,  et~al., 2011, \mn@doi [\prl] {10.1103/PhysRevLett.106.241101},
  \href {https://ui.adsabs.harvard.edu/abs/2011PhRvL.106x1101A} {106, 241101}

\bibitem[\protect\citeauthoryear{{Antonini}, {Gieles}  \&
  {Gualandris}}{{Antonini} et~al.}{2019}]{antonini19}
{Antonini} F.,  {Gieles} M.,   {Gualandris} A.,  2019, \mn@doi [\mnras]
  {10.1093/mnras/stz1149}, \href
  {https://ui.adsabs.harvard.edu/abs/2019MNRAS.486.5008A} {486, 5008}

\bibitem[\protect\citeauthoryear{{Ashton}, {Ackley}, {Hernandez}  \&
  {Piotrzkowski}}{{Ashton} et~al.}{2021}]{ashton21}
{Ashton} G.,  {Ackley} K.,  {Hernandez} I.~M.,   {Piotrzkowski} B.,  2021,
  \mn@doi [Classical and Quantum Gravity] {10.1088/1361-6382/ac33bb}, \href
  {https://ui.adsabs.harvard.edu/abs/2021CQGra..38w5004A} {38, 235004}

\bibitem[\protect\citeauthoryear{{Assef} et~al.,}{{Assef}
  et~al.}{2013}]{assef13}
{Assef} R.~J.,  et~al., 2013, \mn@doi [\apj] {10.1088/0004-637X/772/1/26},
  \href {https://ui.adsabs.harvard.edu/abs/2013ApJ...772...26A} {772, 26}

\bibitem[\protect\citeauthoryear{{Astropy Collaboration} et~al.,}{{Astropy
  Collaboration} et~al.}{2013}]{astropy13}
{Astropy Collaboration} et~al., 2013, \mn@doi [\aap]
  {10.1051/0004-6361/201322068}, \href
  {https://ui.adsabs.harvard.edu/abs/2013A&A...558A..33A} {558, A33}

\bibitem[\protect\citeauthoryear{{Astropy Collaboration} et~al.,}{{Astropy
  Collaboration} et~al.}{2018}]{astropy18}
{Astropy Collaboration} et~al., 2018, \mn@doi [\aj] {10.3847/1538-3881/aabc4f},
  \href {https://ui.adsabs.harvard.edu/abs/2018AJ....156..123A} {156, 123}

\bibitem[\protect\citeauthoryear{{Barrera} \& {Bartos}}{{Barrera} \&
  {Bartos}}{2022}]{barrera22}
{Barrera} O.,  {Bartos} I.,  2022, \mn@doi [\apjl] {10.3847/2041-8213/ac5f47},
  \href {https://ui.adsabs.harvard.edu/abs/2022ApJ...929L...1B} {929, L1}

\bibitem[\protect\citeauthoryear{{Bartos}}{{Bartos}}{2016}]{bartos16}
{Bartos} I.,  2016, in American Astronomical Society Meeting Abstracts \#228.
  p. 208.03

\bibitem[\protect\citeauthoryear{{Bartos}, {Haiman}, {Marka}, {Metzger},
  {Stone}  \& {Marka}}{{Bartos} et~al.}{2017}]{bartos17}
{Bartos} I.,  {Haiman} Z.,  {Marka} Z.,  {Metzger} B.~D.,  {Stone} N.~C.,
  {Marka} S.,  2017, \mn@doi [Nature Communications]
  {10.1038/s41467-017-00851-7}, \href
  {https://ui.adsabs.harvard.edu/abs/2017NatCo...8..831B} {8, 831}

\bibitem[\protect\citeauthoryear{{Belczynski} et~al.,}{{Belczynski}
  et~al.}{2016}]{belczynski16}
{Belczynski} K.,  et~al., 2016, \mn@doi [\aap] {10.1051/0004-6361/201628980},
  \href {https://ui.adsabs.harvard.edu/abs/2016A&A...594A..97B} {594, A97}

\bibitem[\protect\citeauthoryear{{Belczynski}, {Doctor}, {Zevin}, {Olejak},
  {Banerje}  \& {Chattopadhyay}}{{Belczynski} et~al.}{2022}]{belczynski22}
{Belczynski} K.,  {Doctor} Z.,  {Zevin} M.,  {Olejak} A.,  {Banerje} S.,
  {Chattopadhyay} D.,  2022, \mn@doi [\apj] {10.3847/1538-4357/ac8167}, \href
  {https://ui.adsabs.harvard.edu/abs/2022ApJ...935..126B} {935, 126}

\bibitem[\protect\citeauthoryear{{Bellovary}, {Mac Low}, {McKernan}  \&
  {Ford}}{{Bellovary} et~al.}{2016}]{bellovary16}
{Bellovary} J.~M.,  {Mac Low} M.-M.,  {McKernan} B.,   {Ford} K.~E.~S.,  2016,
  \mn@doi [\apjl] {10.3847/2041-8205/819/2/L17}, \href
  {https://ui.adsabs.harvard.edu/abs/2016ApJ...819L..17B} {819, L17}

\bibitem[\protect\citeauthoryear{{Blanton} et~al.,}{{Blanton}
  et~al.}{2017}]{blanton17}
{Blanton} M.~R.,  et~al., 2017, \mn@doi [\aj] {10.3847/1538-3881/aa7567}, \href
  {https://ui.adsabs.harvard.edu/abs/2017AJ....154...28B} {154, 28}

\bibitem[\protect\citeauthoryear{{Braun}, {Dumm}, {De Palma}, {Finley}, {Karle}
   \& {Montaruli}}{{Braun} et~al.}{2008}]{braun08}
{Braun} J.,  {Dumm} J.,  {De Palma} F.,  {Finley} C.,  {Karle} A.,
  {Montaruli} T.,  2008, \mn@doi [Astroparticle Physics]
  {10.1016/j.astropartphys.2008.02.007}, \href
  {https://ui.adsabs.harvard.edu/abs/2008APh....29..299B} {29, 299}

\bibitem[\protect\citeauthoryear{{Corley} et~al.,}{{Corley}
  et~al.}{2019}]{corley19}
{Corley} K.~R.,  et~al., 2019, \mn@doi [\mnras] {10.1093/mnras/stz2072}, \href
  {https://ui.adsabs.harvard.edu/abs/2019MNRAS.488.4459C} {488, 4459}

\bibitem[\protect\citeauthoryear{{Costa}, {Bressan}, {Mapelli}, {Marigo},
  {Iorio}  \& {Spera}}{{Costa} et~al.}{2021}]{costa21}
{Costa} G.,  {Bressan} A.,  {Mapelli} M.,  {Marigo} P.,  {Iorio} G.,   {Spera}
  M.,  2021, \mn@doi [\mnras] {10.1093/mnras/staa3916}, \href
  {https://ui.adsabs.harvard.edu/abs/2021MNRAS.501.4514C} {501, 4514}

\bibitem[\protect\citeauthoryear{{DeLaurentiis}, {Epstein-Martin}  \&
  {Haiman}}{{DeLaurentiis} et~al.}{2022}]{delaurentiis22}
{DeLaurentiis} S.,  {Epstein-Martin} M.,   {Haiman} Z.,  2022, \mn@doi [arXiv
  e-prints] {10.48550/arXiv.2212.02650}, \href
  {https://ui.adsabs.harvard.edu/abs/2022arXiv221202650D} {p. arXiv:2212.02650}

\bibitem[\protect\citeauthoryear{{Fabj}, {Nasim}, {Caban}, {Ford}, {McKernan}
  \& {Bellovary}}{{Fabj} et~al.}{2020}]{fabj20}
{Fabj} G.,  {Nasim} S.~S.,  {Caban} F.,  {Ford} K.~E.~S.,  {McKernan} B.,
  {Bellovary} J.~M.,  2020, \mn@doi [\mnras] {10.1093/mnras/staa3004}, \href
  {https://ui.adsabs.harvard.edu/abs/2020MNRAS.499.2608F} {499, 2608}

\bibitem[\protect\citeauthoryear{{Fishbach}, {Kimball}  \&
  {Kalogera}}{{Fishbach} et~al.}{2022}]{fishbach22}
{Fishbach} M.,  {Kimball} C.,   {Kalogera} V.,  2022, \mn@doi [\apjl]
  {10.3847/2041-8213/ac86c4}, \href
  {https://ui.adsabs.harvard.edu/abs/2022ApJ...935L..26F} {935, L26}

\bibitem[\protect\citeauthoryear{{Flesch}}{{Flesch}}{2021}]{flesch21}
{Flesch} E.~W.,  2021, VizieR Online Data Catalog, \href
  {https://ui.adsabs.harvard.edu/abs/2021yCat.7290....0F} {p. VII/290}

\bibitem[\protect\citeauthoryear{{Ford} \& {McKernan}}{{Ford} \&
  {McKernan}}{2022}]{ford22}
{Ford} K.~E.~S.,  {McKernan} B.,  2022, \mn@doi [\mnras]
  {10.1093/mnras/stac2861}, \href
  {https://ui.adsabs.harvard.edu/abs/2022MNRAS.517.5827F} {517, 5827}

\bibitem[\protect\citeauthoryear{{Gayathri}, {Yang}, {Tagawa}, {Haiman}  \&
  {Bartos}}{{Gayathri} et~al.}{2021}]{gayathri21}
{Gayathri} V.,  {Yang} Y.,  {Tagawa} H.,  {Haiman} Z.,   {Bartos} I.,  2021,
  \mn@doi [\apjl] {10.3847/2041-8213/ac2cc1}, \href
  {https://ui.adsabs.harvard.edu/abs/2021ApJ...920L..42G} {920, L42}

\bibitem[\protect\citeauthoryear{{Gayathri}, {Wysocki}, {Yang}, {Shaughnessy},
  {Haiman}, {Tagawa}  \& {Bartos}}{{Gayathri} et~al.}{2023}]{gayathri23}
{Gayathri} V.,  {Wysocki} D.,  {Yang} Y.,  {Shaughnessy} R.~O.,  {Haiman} Z.,
  {Tagawa} H.,   {Bartos} I.,  2023, \mn@doi [arXiv e-prints]
  {10.48550/arXiv.2301.04187}, \href
  {https://ui.adsabs.harvard.edu/abs/2023arXiv230104187G} {p. arXiv:2301.04187}

\bibitem[\protect\citeauthoryear{{Gerosa} \& {Berti}}{{Gerosa} \&
  {Berti}}{2017}]{gerosa17}
{Gerosa} D.,  {Berti} E.,  2017, \mn@doi [\prd] {10.1103/PhysRevD.95.124046},
  \href {https://ui.adsabs.harvard.edu/abs/2017PhRvD..95l4046G} {95, 124046}

\bibitem[\protect\citeauthoryear{{Gerosa} \& {Berti}}{{Gerosa} \&
  {Berti}}{2019}]{gerosa19}
{Gerosa} D.,  {Berti} E.,  2019, \mn@doi [\prd] {10.1103/PhysRevD.100.041301},
  \href {https://ui.adsabs.harvard.edu/abs/2019PhRvD.100d1301G} {100, 041301}

\bibitem[\protect\citeauthoryear{{Gerosa} \& {Fishbach}}{{Gerosa} \&
  {Fishbach}}{2021}]{gerosa21}
{Gerosa} D.,  {Fishbach} M.,  2021, \mn@doi [Nature Astronomy]
  {10.1038/s41550-021-01398-w}, \href
  {https://ui.adsabs.harvard.edu/abs/2021NatAs...5..749G} {5, 749}

\bibitem[\protect\citeauthoryear{{Graham} et~al.,}{{Graham}
  et~al.}{2020}]{graham20}
{Graham} M.~J.,  et~al., 2020, \mn@doi [\prl] {10.1103/PhysRevLett.124.251102},
  \href {https://ui.adsabs.harvard.edu/abs/2020PhRvL.124y1102G} {124, 251102}

\bibitem[\protect\citeauthoryear{{Graham} et~al.,}{{Graham}
  et~al.}{2023}]{graham23}
{Graham} M.~J.,  et~al., 2023, \mn@doi [\apj] {10.3847/1538-4357/aca480}, \href
  {https://ui.adsabs.harvard.edu/abs/2023ApJ...942...99G} {942, 99}

\bibitem[\protect\citeauthoryear{{Grishin}, {Gilbaum}  \& {Stone}}{{Grishin}
  et~al.}{2023}]{grishin23}
{Grishin} E.,  {Gilbaum} S.,   {Stone} N.~C.,  2023, \mn@doi [arXiv e-prints]
  {10.48550/arXiv.2307.07546}, \href
  {https://ui.adsabs.harvard.edu/abs/2023arXiv230707546G} {p. arXiv:2307.07546}

\bibitem[\protect\citeauthoryear{{Harris} et~al.,}{{Harris}
  et~al.}{2020}]{harris20}
{Harris} C.~R.,  et~al., 2020, \mn@doi [\nat] {10.1038/s41586-020-2649-2},
  \href {https://ui.adsabs.harvard.edu/abs/2020Natur.585..357H} {585, 357}

\bibitem[\protect\citeauthoryear{{Heger} \& {Woosley}}{{Heger} \&
  {Woosley}}{2002}]{heger02}
{Heger} A.,  {Woosley} S.~E.,  2002, \mn@doi [\apj] {10.1086/338487}, \href
  {https://ui.adsabs.harvard.edu/abs/2002ApJ...567..532H} {567, 532}

\bibitem[\protect\citeauthoryear{{Hills} \& {Fullerton}}{{Hills} \&
  {Fullerton}}{1980}]{hills80}
{Hills} J.~G.,  {Fullerton} L.~W.,  1980, \mn@doi [\aj] {10.1086/112798}, \href
  {https://ui.adsabs.harvard.edu/abs/1980AJ.....85.1281H} {85, 1281}

\bibitem[\protect\citeauthoryear{{Hopkins}, {Richards}  \&
  {Hernquist}}{{Hopkins} et~al.}{2007}]{hopkins07}
{Hopkins} P.~F.,  {Richards} G.~T.,   {Hernquist} L.,  2007, \mn@doi [\apj]
  {10.1086/509629}, \href
  {https://ui.adsabs.harvard.edu/abs/2007ApJ...654..731H} {654, 731}

\bibitem[\protect\citeauthoryear{{Hunter}}{{Hunter}}{2007}]{hunter07}
{Hunter} J.~D.,  2007, \mn@doi [Computing in Science and Engineering]
  {10.1109/MCSE.2007.55}, \href
  {https://ui.adsabs.harvard.edu/abs/2007CSE.....9...90H} {9, 90}

\bibitem[\protect\citeauthoryear{{Husa}, {Khan}, {Hannam}, {P{\"u}rrer},
  {Ohme}, {Forteza}  \& {Boh{\'e}}}{{Husa} et~al.}{2016}]{husa16}
{Husa} S.,  {Khan} S.,  {Hannam} M.,  {P{\"u}rrer} M.,  {Ohme} F.,  {Forteza}
  X.~J.,   {Boh{\'e}} A.,  2016, \mn@doi [\prd] {10.1103/PhysRevD.93.044006},
  \href {https://ui.adsabs.harvard.edu/abs/2016PhRvD..93d4006H} {93, 044006}

\bibitem[\protect\citeauthoryear{{Karathanasis}, {Mukherjee}  \&
  {Mastrogiovanni}}{{Karathanasis} et~al.}{2022}]{karathanasis22}
{Karathanasis} C.,  {Mukherjee} S.,   {Mastrogiovanni} S.,  2022, \mn@doi
  [arXiv e-prints] {10.48550/arXiv.2204.13495}, \href
  {https://ui.adsabs.harvard.edu/abs/2022arXiv220413495K} {p. arXiv:2204.13495}

\bibitem[\protect\citeauthoryear{{Khan}, {Husa}, {Hannam}, {Ohme},
  {P{\"u}rrer}, {Forteza}  \& {Boh{\'e}}}{{Khan} et~al.}{2016}]{khan16}
{Khan} S.,  {Husa} S.,  {Hannam} M.,  {Ohme} F.,  {P{\"u}rrer} M.,  {Forteza}
  X.~J.,   {Boh{\'e}} A.,  2016, \mn@doi [\prd] {10.1103/PhysRevD.93.044007},
  \href {https://ui.adsabs.harvard.edu/abs/2016PhRvD..93d4007K} {93, 044007}

\bibitem[\protect\citeauthoryear{{Kritos}, {Berti}  \& {Silk}}{{Kritos}
  et~al.}{2022}]{kritos22}
{Kritos} K.,  {Berti} E.,   {Silk} J.,  2022, \mn@doi [arXiv e-prints]
  {10.48550/arXiv.2212.06845}, \href
  {https://ui.adsabs.harvard.edu/abs/2022arXiv221206845K} {p. arXiv:2212.06845}

\bibitem[\protect\citeauthoryear{{LIGO Scientific Collaboration} et~al.,}{{LIGO
  Scientific Collaboration} et~al.}{2015}]{ligo15}
{LIGO Scientific Collaboration} et~al., 2015, \mn@doi [Classical and Quantum
  Gravity] {10.1088/0264-9381/32/7/074001}, \href
  {https://ui.adsabs.harvard.edu/abs/2015CQGra..32g4001L} {32, 074001}

\bibitem[\protect\citeauthoryear{{Lamontagne}, {Demers}, {Wesemael}, {Fontaine}
   \& {Irwin}}{{Lamontagne} et~al.}{2000}]{lamontagne00}
{Lamontagne} R.,  {Demers} S.,  {Wesemael} F.,  {Fontaine} G.,   {Irwin} M.~J.,
   2000, \mn@doi [\aj] {10.1086/301181}, \href
  {https://ui.adsabs.harvard.edu/abs/2000AJ....119..241L} {119, 241}

\bibitem[\protect\citeauthoryear{{Li} \& {Lai}}{{Li} \& {Lai}}{2022a}]{li22b}
{Li} R.,  {Lai} D.,  2022a, \mn@doi [arXiv e-prints]
  {10.48550/arXiv.2207.01125}, \href
  {https://ui.adsabs.harvard.edu/abs/2022arXiv220701125L} {p. arXiv:2207.01125}

\bibitem[\protect\citeauthoryear{{Li} \& {Lai}}{{Li} \& {Lai}}{2022b}]{li22}
{Li} R.,  {Lai} D.,  2022b, \mn@doi [\mnras] {10.1093/mnras/stac2577}, \href
  {https://ui.adsabs.harvard.edu/abs/2022MNRAS.517.1602L} {517, 1602}

\bibitem[\protect\citeauthoryear{{Li}, {Lin}  \& {Yuan}}{{Li}
  et~al.}{2022}]{lilin22}
{Li} G.-P.,  {Lin} D.-B.,   {Yuan} Y.,  2022, \mn@doi [arXiv e-prints]
  {10.48550/arXiv.2211.11150}, \href
  {https://ui.adsabs.harvard.edu/abs/2022arXiv221111150L} {p. arXiv:2211.11150}

\bibitem[\protect\citeauthoryear{{Liu}, {Liu}, {Dong}, {Zhou}, {Wang}, {Lu}  \&
  {Yuan}}{{Liu} et~al.}{2019}]{liu19}
{Liu} H.-Y.,  {Liu} W.-J.,  {Dong} X.-B.,  {Zhou} H.,  {Wang} T.,  {Lu} H.,
  {Yuan} W.,  2019, \mn@doi [\apjs] {10.3847/1538-4365/ab298b}, \href
  {https://ui.adsabs.harvard.edu/abs/2019ApJS..243...21L} {243, 21}

\bibitem[\protect\citeauthoryear{{Loeb}}{{Loeb}}{2016}]{loeb16}
{Loeb} A.,  2016, \mn@doi [\apjl] {10.3847/2041-8205/819/2/L21}, \href
  {https://ui.adsabs.harvard.edu/abs/2016ApJ...819L..21L} {819, L21}

\bibitem[\protect\citeauthoryear{{Lyke} et~al.,}{{Lyke} et~al.}{2020}]{lyke20}
{Lyke} B.~W.,  et~al., 2020, \mn@doi [\apjs] {10.3847/1538-4365/aba623}, \href
  {https://ui.adsabs.harvard.edu/abs/2020ApJS..250....8L} {250, 8}

\bibitem[\protect\citeauthoryear{{Mahapatra}, {Gupta}, {Favata}, {Arun}  \&
  {Sathyaprakash}}{{Mahapatra} et~al.}{2022}]{mahapatra22}
{Mahapatra} P.,  {Gupta} A.,  {Favata} M.,  {Arun} K.~G.,   {Sathyaprakash}
  B.~S.,  2022, \mn@doi [arXiv e-prints] {10.48550/arXiv.2209.05766}, \href
  {https://ui.adsabs.harvard.edu/abs/2022arXiv220905766M} {p. arXiv:2209.05766}

\bibitem[\protect\citeauthoryear{{Mapelli}}{{Mapelli}}{2021}]{mapelli21}
{Mapelli} M.,  2021, arXiv e-prints, \href
  {https://ui.adsabs.harvard.edu/abs/2021arXiv210600699M} {p. arXiv:2106.00699}

\bibitem[\protect\citeauthoryear{{Masci}, {Cutri}, {Francis}, {Nelson},
  {Huchra}, {Heath Jones}, {Colless}  \& {Saunders}}{{Masci}
  et~al.}{2010}]{masci10}
{Masci} F.~J.,  {Cutri} R.~M.,  {Francis} P.~J.,  {Nelson} B.~O.,  {Huchra}
  J.~P.,  {Heath Jones} D.,  {Colless} M.,   {Saunders} W.,  2010, \mn@doi
  [\pasa] {10.1071/AS10001}, \href
  {https://ui.adsabs.harvard.edu/abs/2010PASA...27..302M} {27, 302}

\bibitem[\protect\citeauthoryear{{Mauch} \& {Sadler}}{{Mauch} \&
  {Sadler}}{2007}]{mauch07}
{Mauch} T.,  {Sadler} E.~M.,  2007, VizieR Online Data Catalog, \href
  {https://ui.adsabs.harvard.edu/abs/2007yCat..83750931M} {p. J/MNRAS/375/931}

\bibitem[\protect\citeauthoryear{{McKernan}, {Ford}, {Lyra}, {Perets}, {Winter}
   \& {Yaqoob}}{{McKernan} et~al.}{2011}]{mckernan12}
{McKernan} B.,  {Ford} K.~E.~S.,  {Lyra} W.,  {Perets} H.~B.,  {Winter} L.~M.,
   {Yaqoob} T.,  2011, \mn@doi [\mnras] {10.1111/j.1745-3933.2011.01132.x},
  \href {https://ui.adsabs.harvard.edu/abs/2011MNRAS.417L.103M} {417, L103}

\bibitem[\protect\citeauthoryear{{McKernan}, {Ford}, {Lyra}  \&
  {Perets}}{{McKernan} et~al.}{2012}]{mckernan12b}
{McKernan} B.,  {Ford} K.~E.~S.,  {Lyra} W.,   {Perets} H.~B.,  2012, \mn@doi
  [\mnras] {10.1111/j.1365-2966.2012.21486.x}, \href
  {https://ui.adsabs.harvard.edu/abs/2012MNRAS.425..460M} {425, 460}

\bibitem[\protect\citeauthoryear{{McKernan} et~al.,}{{McKernan}
  et~al.}{2019}]{mckernan19}
{McKernan} B.,  et~al., 2019, \mn@doi [\apjl] {10.3847/2041-8213/ab4886}, \href
  {https://ui.adsabs.harvard.edu/abs/2019ApJ...884L..50M} {884, L50}

\bibitem[\protect\citeauthoryear{{McKernan}, {Ford}, {O'Shaugnessy}  \&
  {Wysocki}}{{McKernan} et~al.}{2020}]{mckernan20}
{McKernan} B.,  {Ford} K.~E.~S.,  {O'Shaugnessy} R.,   {Wysocki} D.,  2020,
  \mn@doi [\mnras] {10.1093/mnras/staa740}, \href
  {https://ui.adsabs.harvard.edu/abs/2020MNRAS.494.1203M} {494, 1203}

\bibitem[\protect\citeauthoryear{{McKernan}, {Ford}, {Cantiello}, {Graham},
  {Jermyn}, {Leigh}, {Ryu}  \& {Stern}}{{McKernan} et~al.}{2022}]{mckernan22}
{McKernan} B.,  {Ford} K.~E.~S.,  {Cantiello} M.,  {Graham} M.,  {Jermyn}
  A.~S.,  {Leigh} N.~W.~C.,  {Ryu} T.,   {Stern} D.,  2022, \mn@doi [\mnras]
  {10.1093/mnras/stac1310}, \href
  {https://ui.adsabs.harvard.edu/abs/2022MNRAS.514.4102M} {514, 4102}

\bibitem[\protect\citeauthoryear{{Monroe}, {Prochaska}, {Tejos}, {Worseck},
  {Hennawi}, {Schmidt}, {Tumlinson}  \& {Shen}}{{Monroe}
  et~al.}{2016}]{monroe16}
{Monroe} T.~R.,  {Prochaska} J.~X.,  {Tejos} N.,  {Worseck} G.,  {Hennawi}
  J.~F.,  {Schmidt} T.,  {Tumlinson} J.,   {Shen} Y.,  2016, \mn@doi [\aj]
  {10.3847/0004-6256/152/1/25}, \href
  {https://ui.adsabs.harvard.edu/abs/2016AJ....152...25M} {152, 25}

\bibitem[\protect\citeauthoryear{{Palenzuela}, {Lehner}  \&
  {Yoshida}}{{Palenzuela} et~al.}{2010}]{palenzuela10}
{Palenzuela} C.,  {Lehner} L.,   {Yoshida} S.,  2010, \mn@doi [\prd]
  {10.1103/PhysRevD.81.084007}, \href
  {https://ui.adsabs.harvard.edu/abs/2010PhRvD..81h4007P} {81, 084007}

\bibitem[\protect\citeauthoryear{{Paturel}, {Petit}, {Prugniel}, {Theureau},
  {Rousseau}, {Brouty}, {Dubois}  \& {Cambr{\'e}sy}}{{Paturel}
  et~al.}{2003}]{paturel03}
{Paturel} G.,  {Petit} C.,  {Prugniel} P.,  {Theureau} G.,  {Rousseau} J.,
  {Brouty} M.,  {Dubois} P.,   {Cambr{\'e}sy} L.,  2003, \mn@doi [\aap]
  {10.1051/0004-6361:20031411}, \href
  {https://ui.adsabs.harvard.edu/abs/2003A&A...412...45P} {412, 45}

\bibitem[\protect\citeauthoryear{{Peng} \& {Chen}}{{Peng} \&
  {Chen}}{2021}]{peng21}
{Peng} P.,  {Chen} X.,  2021, \mn@doi [\mnras] {10.1093/mnras/stab1419}, \href
  {https://ui.adsabs.harvard.edu/abs/2021MNRAS.505.1324P} {505, 1324}

\bibitem[\protect\citeauthoryear{{Petrov} et~al.,}{{Petrov}
  et~al.}{2022}]{petrov22}
{Petrov} P.,  et~al., 2022, \mn@doi [\apj] {10.3847/1538-4357/ac366d}, \href
  {https://ui.adsabs.harvard.edu/abs/2022ApJ...924...54P} {924, 54}

\bibitem[\protect\citeauthoryear{{Planck Collaboration} et~al.,}{{Planck
  Collaboration} et~al.}{2016}]{planck15}
{Planck Collaboration} et~al., 2016, \mn@doi [\aap]
  {10.1051/0004-6361/201525830}, \href
  {https://ui.adsabs.harvard.edu/abs/2016A&A...594A..13P} {594, A13}

\bibitem[\protect\citeauthoryear{{Qin} et~al.,}{{Qin} et~al.}{2022}]{qin22}
{Qin} Y.,  et~al., 2022, \mn@doi [\apj] {10.3847/1538-4357/aca40c}, \href
  {https://ui.adsabs.harvard.edu/abs/2022ApJ...941..179Q} {941, 179}

\bibitem[\protect\citeauthoryear{{Richards} et~al.,}{{Richards}
  et~al.}{2002}]{richards02}
{Richards} G.~T.,  et~al., 2002, \mn@doi [\aj] {10.1086/340187}, \href
  {https://ui.adsabs.harvard.edu/abs/2002AJ....123.2945R} {123, 2945}

\bibitem[\protect\citeauthoryear{{Rodriguez} \& {Loeb}}{{Rodriguez} \&
  {Loeb}}{2018}]{rodriguez18}
{Rodriguez} C.~L.,  {Loeb} A.,  2018, \mn@doi [\apjl]
  {10.3847/2041-8213/aae377}, \href
  {https://ui.adsabs.harvard.edu/abs/2018ApJ...866L...5R} {866, L5}

\bibitem[\protect\citeauthoryear{{Rodriguez}, {Chatterjee}  \&
  {Rasio}}{{Rodriguez} et~al.}{2016}]{rodriguez16}
{Rodriguez} C.~L.,  {Chatterjee} S.,   {Rasio} F.~A.,  2016, \mn@doi [\prd]
  {10.1103/PhysRevD.93.084029}, \href
  {https://ui.adsabs.harvard.edu/abs/2016PhRvD..93h4029R} {93, 084029}

\bibitem[\protect\citeauthoryear{{Rodriguez}, {Kremer}, {Chatterjee},
  {Fragione}, {Loeb}, {Rasio}, {Weatherford}  \& {Ye}}{{Rodriguez}
  et~al.}{2021}]{rodriguez21}
{Rodriguez} C.~L.,  {Kremer} K.,  {Chatterjee} S.,  {Fragione} G.,  {Loeb} A.,
  {Rasio} F.~A.,  {Weatherford} N.~C.,   {Ye} C.~S.,  2021, \mn@doi [Research
  Notes of the American Astronomical Society] {10.3847/2515-5172/abdf54}, \href
  {https://ui.adsabs.harvard.edu/abs/2021RNAAS...5...19R} {5, 19}

\bibitem[\protect\citeauthoryear{{Romero-Shaw}, {Lasky}  \&
  {Thrane}}{{Romero-Shaw} et~al.}{2021}]{romeroshaw21}
{Romero-Shaw} I.,  {Lasky} P.~D.,   {Thrane} E.,  2021, \mn@doi [\apjl]
  {10.3847/2041-8213/ac3138}, \href
  {https://ui.adsabs.harvard.edu/abs/2021ApJ...921L..31R} {921, L31}

\bibitem[\protect\citeauthoryear{{Romero-Shaw}, {Lasky}  \&
  {Thrane}}{{Romero-Shaw} et~al.}{2022}]{romeroshaw22}
{Romero-Shaw} I.,  {Lasky} P.~D.,   {Thrane} E.,  2022, \mn@doi [\apj]
  {10.3847/1538-4357/ac9798}, \href
  {https://ui.adsabs.harvard.edu/abs/2022ApJ...940..171R} {940, 171}

\bibitem[\protect\citeauthoryear{{Rowan}, {Boekholt}, {Kocsis}  \&
  {Haiman}}{{Rowan} et~al.}{2022}]{rowan22}
{Rowan} C.,  {Boekholt} T.,  {Kocsis} B.,   {Haiman} Z.,  2022, \mn@doi [arXiv
  e-prints] {10.48550/arXiv.2212.06133}, \href
  {https://ui.adsabs.harvard.edu/abs/2022arXiv221206133R} {p. arXiv:2212.06133}

\bibitem[\protect\citeauthoryear{{Samsing} et~al.,}{{Samsing}
  et~al.}{2022}]{samsing22}
{Samsing} J.,  et~al., 2022, \mn@doi [\nat] {10.1038/s41586-021-04333-1}, \href
  {https://ui.adsabs.harvard.edu/abs/2022Natur.603..237S} {603, 237}

\bibitem[\protect\citeauthoryear{{Schlafly}, {Meisner}  \& {Green}}{{Schlafly}
  et~al.}{2019}]{schlafly19}
{Schlafly} E.~F.,  {Meisner} A.~M.,   {Green} G.~M.,  2019, \mn@doi [\apjs]
  {10.3847/1538-4365/aafbea}, \href
  {https://ui.adsabs.harvard.edu/abs/2019ApJS..240...30S} {240, 30}

\bibitem[\protect\citeauthoryear{{Singer} \& {Price}}{{Singer} \&
  {Price}}{2016}]{singer16}
{Singer} L.~P.,  {Price} L.~R.,  2016, \mn@doi [\prd]
  {10.1103/PhysRevD.93.024013}, \href
  {https://ui.adsabs.harvard.edu/abs/2016PhRvD..93b4013S} {93, 024013}

\bibitem[\protect\citeauthoryear{{Stern} et~al.,}{{Stern}
  et~al.}{2012}]{stern12}
{Stern} D.,  et~al., 2012, \mn@doi [\apj] {10.1088/0004-637X/753/1/30}, \href
  {https://ui.adsabs.harvard.edu/abs/2012ApJ...753...30S} {753, 30}

\bibitem[\protect\citeauthoryear{{Stevenson} \& {Clarke}}{{Stevenson} \&
  {Clarke}}{2022}]{stevenson22}
{Stevenson} S.,  {Clarke} T.~A.,  2022, \mn@doi [\mnras]
  {10.1093/mnras/stac2936}, \href
  {https://ui.adsabs.harvard.edu/abs/2022MNRAS.517.4034S} {517, 4034}

\bibitem[\protect\citeauthoryear{{Stone}, {Metzger}  \& {Haiman}}{{Stone}
  et~al.}{2017}]{stone17}
{Stone} N.~C.,  {Metzger} B.~D.,   {Haiman} Z.,  2017, \mn@doi [\mnras]
  {10.1093/mnras/stw2260}, \href
  {https://ui.adsabs.harvard.edu/abs/2017MNRAS.464..946S} {464, 946}

\bibitem[\protect\citeauthoryear{{Strauss} et~al.,}{{Strauss}
  et~al.}{2002}]{strauss02}
{Strauss} M.~A.,  et~al., 2002, \mn@doi [\aj] {10.1086/342343}, \href
  {https://ui.adsabs.harvard.edu/abs/2002AJ....124.1810S} {124, 1810}

\bibitem[\protect\citeauthoryear{{Tagawa}, {Haiman}  \& {Kocsis}}{{Tagawa}
  et~al.}{2020}]{tagawa20}
{Tagawa} H.,  {Haiman} Z.,   {Kocsis} B.,  2020, \mn@doi [\apj]
  {10.3847/1538-4357/ab9b8c}, \href
  {https://ui.adsabs.harvard.edu/abs/2020ApJ...898...25T} {898, 25}

\bibitem[\protect\citeauthoryear{{Tagawa}, {Haiman}, {Bartos}, {Kocsis}  \&
  {Omukai}}{{Tagawa} et~al.}{2021}]{tagawa21}
{Tagawa} H.,  {Haiman} Z.,  {Bartos} I.,  {Kocsis} B.,   {Omukai} K.,  2021,
  \mn@doi [\mnras] {10.1093/mnras/stab2315}, \href
  {https://ui.adsabs.harvard.edu/abs/2021MNRAS.507.3362T} {507, 3362}

\bibitem[\protect\citeauthoryear{{Tanikawa}, {Susa}, {Yoshida}, {Trani}  \&
  {Kinugawa}}{{Tanikawa} et~al.}{2021}]{tanikawa21}
{Tanikawa} A.,  {Susa} H.,  {Yoshida} T.,  {Trani} A.~A.,   {Kinugawa} T.,
  2021, \mn@doi [\apj] {10.3847/1538-4357/abe40d}, \href
  {https://ui.adsabs.harvard.edu/abs/2021ApJ...910...30T} {910, 30}

\bibitem[\protect\citeauthoryear{{Tesch} \& {Engels}}{{Tesch} \&
  {Engels}}{2000}]{tesch00}
{Tesch} F.,  {Engels} D.,  2000, \mn@doi [\mnras]
  {10.1046/j.1365-8711.2000.03264.x}, \href
  {https://ui.adsabs.harvard.edu/abs/2000MNRAS.313..377T} {313, 377}

\bibitem[\protect\citeauthoryear{{The LIGO Scientific Collaboration}
  et~al.,}{{The LIGO Scientific Collaboration} et~al.}{2021a}]{ligo21}
{The LIGO Scientific Collaboration} et~al., 2021a, \mn@doi [arXiv e-prints]
  {10.48550/arXiv.2111.03606}, \href
  {https://ui.adsabs.harvard.edu/abs/2021arXiv211103606T} {p. arXiv:2111.03606}

\bibitem[\protect\citeauthoryear{{The LIGO Scientific Collaboration}
  et~al.,}{{The LIGO Scientific Collaboration} et~al.}{2021b}]{ligo21pop}
{The LIGO Scientific Collaboration} et~al., 2021b, \mn@doi [arXiv e-prints]
  {10.48550/arXiv.2111.03634}, \href
  {https://ui.adsabs.harvard.edu/abs/2021arXiv211103634T} {p. arXiv:2111.03634}

\bibitem[\protect\citeauthoryear{{Vajpeyi}, {Thrane}, {Smith}, {McKernan}  \&
  {Saavik Ford}}{{Vajpeyi} et~al.}{2022}]{vajpeyi22}
{Vajpeyi} A.,  {Thrane} E.,  {Smith} R.,  {McKernan} B.,   {Saavik Ford} K.~E.,
   2022, \mn@doi [\apj] {10.3847/1538-4357/ac6180}, \href
  {https://ui.adsabs.harvard.edu/abs/2022ApJ...931...82V} {931, 82}

\bibitem[\protect\citeauthoryear{{Veronesi}, {Rossi}, {van Velzen}  \&
  {Buscicchio}}{{Veronesi} et~al.}{2022}]{veronesi22}
{Veronesi} N.,  {Rossi} E.~M.,  {van Velzen} S.,   {Buscicchio} R.,  2022,
  \mn@doi [\mnras] {10.1093/mnras/stac1346}, \href
  {https://ui.adsabs.harvard.edu/abs/2022MNRAS.514.2092V} {514, 2092}

\bibitem[\protect\citeauthoryear{{Virtanen} et~al.,}{{Virtanen}
  et~al.}{2020}]{virtanen20}
{Virtanen} P.,  et~al., 2020, \mn@doi [Nature Methods]
  {10.1038/s41592-019-0686-2}, \href
  {https://ui.adsabs.harvard.edu/abs/2020NatMe..17..261V} {17, 261}

\bibitem[\protect\citeauthoryear{{Wang}, {Fan}, {Tang}, {Qin}  \& {Wei}}{{Wang}
  et~al.}{2021}]{wang21}
{Wang} Y.-Z.,  {Fan} Y.-Z.,  {Tang} S.-P.,  {Qin} Y.,   {Wei} D.-M.,  2021,
  \mn@doi [arXiv e-prints] {10.48550/arXiv.2110.10838}, \href
  {https://ui.adsabs.harvard.edu/abs/2021arXiv211010838W} {p. arXiv:2110.10838}

\bibitem[\protect\citeauthoryear{{Wang}, {McKernan}, {Ford}, {Perna}, {Leigh}
  \& {Mac Low}}{{Wang} et~al.}{2022}]{wang22}
{Wang} Y.,  {McKernan} B.,  {Ford} K.~E.~S.,  {Perna} R.,  {Leigh} N.,   {Mac
  Low} M.-M.,  2022, in AAS/Division of Dynamical Astronomy Meeting. p. 300.01

\bibitem[\protect\citeauthoryear{{Wei}, {Xu}, {Dong}  \& {Hu}}{{Wei}
  et~al.}{1999}]{wei99}
{Wei} J.~Y.,  {Xu} D.~W.,  {Dong} X.~Y.,   {Hu} J.~Y.,  1999, \mn@doi [\aaps]
  {10.1051/aas:1999514}, \href
  {https://ui.adsabs.harvard.edu/abs/1999A&AS..139..575W} {139, 575}

\bibitem[\protect\citeauthoryear{{Woosley}}{{Woosley}}{2019}]{woosley19}
{Woosley} S.~E.,  2019, \mn@doi [\apj] {10.3847/1538-4357/ab1b41}, \href
  {https://ui.adsabs.harvard.edu/abs/2019ApJ...878...49W} {878, 49}

\bibitem[\protect\citeauthoryear{{Wright} et~al.,}{{Wright}
  et~al.}{2010}]{wright10}
{Wright} E.~L.,  et~al., 2010, \mn@doi [\aj] {10.1088/0004-6256/140/6/1868},
  \href {https://ui.adsabs.harvard.edu/abs/2010AJ....140.1868W} {140, 1868}

\bibitem[\protect\citeauthoryear{{Yang} et~al.,}{{Yang} et~al.}{2019}]{yang19}
{Yang} Y.,  et~al., 2019, \mn@doi [\prl] {10.1103/PhysRevLett.123.181101},
  \href {https://ui.adsabs.harvard.edu/abs/2019PhRvL.123r1101Y} {123, 181101}

\bibitem[\protect\citeauthoryear{{York} et~al.,}{{York} et~al.}{2000}]{york00}
{York} D.~G.,  et~al., 2000, \mn@doi [\aj] {10.1086/301513}, \href
  {https://ui.adsabs.harvard.edu/abs/2000AJ....120.1579Y} {120, 1579}

\bibitem[\protect\citeauthoryear{{Zevin} \& {Bavera}}{{Zevin} \&
  {Bavera}}{2022}]{zevin22}
{Zevin} M.,  {Bavera} S.~S.,  2022, \mn@doi [\apj] {10.3847/1538-4357/ac6f5d},
  \href {https://ui.adsabs.harvard.edu/abs/2022ApJ...933...86Z} {933, 86}

\bibitem[\protect\citeauthoryear{{Ziosi}, {Mapelli}, {Branchesi}  \&
  {Tormen}}{{Ziosi} et~al.}{2014}]{ziosi14}
{Ziosi} B.~M.,  {Mapelli} M.,  {Branchesi} M.,   {Tormen} G.,  2014, \mn@doi
  [\mnras] {10.1093/mnras/stu824}, \href
  {https://ui.adsabs.harvard.edu/abs/2014MNRAS.441.3703Z} {441, 3703}

\bibitem[\protect\citeauthoryear{{de Mink} \& {Mandel}}{{de Mink} \&
  {Mandel}}{2016}]{demink16}
{de Mink} S.~E.,  {Mandel} I.,  2016, \mn@doi [\mnras] {10.1093/mnras/stw1219},
  \href {https://ui.adsabs.harvard.edu/abs/2016MNRAS.460.3545D} {460, 3545}

\makeatother
\end{thebibliography}


\bsp
\label{lastpage}
\end{document}